\documentclass[11pt]{article}
\usepackage{fullpage,citesort,epsfig,psfrag}
\usepackage[normal,small]{caption}
\newcommand{\beq}{\begin{equation}}
\newcommand{\eeq}{\end{equation}}
\newcommand{\bea}{\begin{eqnarray}}
\newcommand{\eea}{\end{eqnarray}}
\newcommand{\intst}{\int_{0}^{p^{+}}d\sigma \int_{\tau_{i}}^{\tau_{f}}
d\tau}

\def\Tr{{\rm Tr}}
\def\to{\rightarrow}
\newcommand{\be}{\begin{equation}}
\newcommand{\ee}{\end{equation}}
\newcommand{\bq}{\begin{eqnarray}}
\newcommand{\eq}{\end{eqnarray}}
\newcommand{\ket}[1]{|#1\rangle}
\newcommand{\bra}[1]{\langle#1|}

\def\ie{{\it i.e.\ }}
\def\m@th{\mathsurround=0pt }
\def\leftrightarrowfill{$\m@th \mathord\leftarrow \mkern-6mu \cleaders\hbox{$\mkern-2mu \mathord- \mkern-2mu$}\hfill
 \mkern-6mu \mathord\rightarrow$}
\def\overleftrightarrow#1{\vbox{\ialign{##\crcr
     \leftrightarrowfill\crcr\noalign{\kern-1pt\nointerlineskip}
     $\hfil\displaystyle{#1}\hfil$\crcr}}}

\begin{document}
\setlength{\captionmargin}{20pt}

\renewcommand{\thefootnote}{\fnsymbol{footnote}}
\begin{titlepage}
\begin{flushright}
LBNL-50597\\UCB-PTH-02/25\\
UFIFT-HEP-02-20\\
hep-th/0206205
\end{flushright}

\vskip 3cm

\begin{center}
\begin{Large}
{\bf A Mean Field Approximation to the \\Worldsheet
Model of Planar $\phi^3$ Field Theory\footnote{
This work was supported in part by the Director, Office of Science,
Office of High Energy and Nuclear Physics, of the U.S. Department
of energy under Contract DE-AC03-76SF00098, in part
by the National Science Foundation Grant PHY-0098840,
in part by the Monell Foundation
and in part by the Department
of Energy under Grant No. DE-FG02-97ER-41029. 
}}
\end{Large}
\vskip 1cm
{\large 
 Korkut Bardakci\footnote{E-mail  address: {\tt kbardakci@lbl.edu}}
}
\vskip 0.5cm
{\it Department of Physics, University of California at Berkeley\\
and\\
Theoretical Physics Group, Lawrence Berkeley National
Laboratory\\ University of California,  Berkeley CA 94720}
\vskip0.5cm 
and
\vskip0.5cm
{\large 
 Charles B. Thorn\footnote{E-mail  address: {\tt thorn@phys.ufl.edu}}
}
\vskip 0.5cm
{\it School of Natural Sciences, Institute for Advanced Study,
Princeton NJ 08540\\
and\\
Institute for Fundamental Theory\\
Department of Physics, University of Florida,
Gainesville FL 32611}


\vskip 1.0cm
\end{center}

\begin{abstract}\noindent
We develop an approximation scheme for our worldsheet
model of the sum of planar diagrams based on mean
field theory. At finite coupling
the mean field equations show a weak coupling solution that
resembles the perturbative diagrams and a strong
coupling solution that seems to represent a tensionless
soup of field quanta. With a certain amount of fine-tuning,
we find a solution of the mean field equations that 
seems to support string formation. 
\end{abstract}

\vfill
\end{titlepage}

\section{Introduction}
Last year we initiated a program \cite{bardakcit}, 
for summing the planar graphs
of large $N_c$ matrix field theories, based on a description of
each planar diagram as a light-cone worldsheet path integral.
Once this worldsheet representation was obtained, we could then
realize the sum over planar graphs by coupling the worldsheet
matter and ghost fields to a two dimensional Ising spin system. Summing over
all spin configurations accomplishes the sum over all
planar diagrams. One can regard the resulting two dimensional
system as a noninteracting string moving in a background
described by the Ising spin system. In this way our proposal
arrives at a result suggested by the Maldacena conjecture \cite{maldacena},
namely the association of a free string theory in a non-trivial
background with the sum of planar diagrams. But instead of
relying on open-string/closed-string duality to provide
equations for the background, we are able to ``read off''
the background directly from the planar diagrams being summed.
Thus we hope our approach represents an attack on the 
problem of large $N_c$ QCD complementary to the well
studied AdS/CFT duality.
 
The worldsheet construction of \cite{bardakcit} was restricted
to $\phi^3$ scalar field theory, but the methods were soon extended
to the important case of pure Yang-Mills field theory in
\cite{thornsheet}. The extension to supersymmetric
field theories remains to be done. But here we return
to the simple $\phi^3$ theory as a useful arena for developing
methods for extracting the physical properties of
the worldsheet system, and hence those of the large $N_c$
limit. In particular we wish to apply the mean field 
approximation to the Ising spin system in order to
get a qualitative understanding of the basic physics described
by the sum of planar graphs. Such an approach has previously been applied
in an attempt to extract
nonperturbative information from lattice string theory
\cite{gilest} in Ref.~\cite{orland}.

At the outset we should take note of the fundamental
instability of $\phi^3$ theory following from the unboundedness
of the cubic potential. As long as the free fields
we perturb around are not tachyonic, this instability does not obstruct
the evaluation of any Feynman graph, nor does it
prevent one from contemplating graph summation
\cite{dalleyk}. Although our real interest is
summing the planar diagrams of a stable non-abelian gauge
theory, we find the simplicity of $\phi^3$ diagrams
particularly valuable as a toy system for developing and testing our
methodology. In particular, we can ask whether our
methods are at all sensitive to the instabilities
we know are there. There is even the possibility that a
metastable vacuum in the finite $N_c$ case becomes
stable in the $N_c\to\infty$ limit \cite{greensiteh}. If so, the
planar sum might give a physically meaningful
answer, and show instability only at higher order in the
$1/N_c$ expansion.

We develop the mean field method in stages. After a brief
review of the worldsheet formalism in Section 2, we begin in Section 3
with a treatment motivated by the formal continuum
limit of the worldsheet model. The qualitative
continuum physics is reviewed and the mean field is
introduced. The basic idea is that the three
worldsheet systems in the construction, namely
the target space variables ${\bf q}$ the ghosts
$b,c$ and the Ising spins $s$, are first exactly solved
in the presence of a homogeneous mean field. Then
the value of the mean field is determined by 
minimizing the sums of the energies of the three
systems. Because the mean field $\phi$ is supposed to represent
a fundamental variable that only takes on the
values $0$ or $1$, there is some
freedom in the way it can enter the dynamics, when $0<\phi<1$,
in the first approximation to the
actions that describe the target space and ghost systems.
Postponing this issue to Section
4, we instead use simple scaling arguments
to fix a reasonable interpolation for the
mean ghost$+$matter energy between its values for $\phi=0$
and $\phi=1$. The resulting mean field equation has a 
solution that supports string formation for a certain
choice of parameters. For this choice there
is a minimum of the energy that is associated
with a finite non-zero string tension.

In Section 4, we develop the mean field approach directly
from the lattice system that was the foundation of
our mapping of planar diagrams to worldsheets. This
treatment is thus more transparently related to the
original field theoretic perturbation theory.
Here we identify the mean field approximation as
a certain saddle-point evaluation of the 
path integral. In this context, the ambiguities mentioned
in the previous paragraphs are simply 
rearrangements of the path integration that are
valid in the exact integral but lead to differences
in the saddle-point approximation. Since the
saddle-point approximation can be systematically
corrected, there is an objective criterion for
choosing the zeroth order action, namely the
one for which the corrections are as small as
possible. We base our choice in this section on
how well it works at zero coupling, where exact
calculations can be made. It turns out that
the consequences of this choice are qualitatively
very similar to the results of Section 3. 
The main difference can be traced to the $\phi^{3/2}$
dependence of the energy in (\ref{smalleps}), as opposed to the linear
dependence on $\phi$ in (\ref{21}). This is due to the ambiguities
in the application of the saddle point method mentioned above,
and it does not lead to any qualitative change in the final
results.
Again there
is a range of parameters, where the coupling
${\hat g}=O(\epsilon)$ where the effective 
string tension is finite and non-zero. Here $\epsilon$ is
a temporary infrared cutoff that must be sent to $0$.
Although the coupling is tending to $0$, this is not
the perturbative regime, which is obtained by expanding
about $g=0$ at fixed $\epsilon$ and lattice cutoffs,
and only removing the cutoffs order by order at the end
of the calculation. For $g/\epsilon\to0$ the effective
string tension becomes infinite, signaling the
return to the perturbative region of weakly coupled
field quanta.

For $g=O(1)$ as $\epsilon\to0$, the effective tension
in our approximation is $0$: the system is
some kind of tensionless soup.
We believe that the unphysical nature of 
this regime is a reflection of the inherent
instability of the initial $\phi^3$ theory. We find
the existence of a regime with
string formation interesting, even though it seems
to require a certain amount of fine-tuning.

Concluding remarks are given in Section 5.
We also include two appendices. In the first we explain the tools
for evaluating the Gaussian path integrals encountered
throughout the article. In the second, we discuss a
simple extension of the mean field description to
slowly varying fields.

\section{Review of the Worldsheet Formalism for Field Theory}
The light-front components of any Minkowski vector
$v^\mu$ will be written $(v^+,v^-,{\bf v})$ or $(v^+,v^-,v^k)$.
Here $v^\pm=(v^0\pm v^3)/\sqrt2$, and the remaining components
label the transverse directions.
The Lorentz invariant
scalar product of two four vectors $v,w$ is written
$v\cdot w={\bf v}\cdot{\bf w}-v^+w^--v^-w^+$. We shall
select $x^+$ to be our quantum evolution parameter, and we
recall that the Hamiltonian conjugate to this time is
$p^-$. A massless on-shell particle thus has the ``energy''
$p^-={\bf p}^2/2p^+$.

The starting point for the worldsheet construction
\cite{bardakcit} is the 
path integral representation of the (imaginary time) light-cone
evolution operator for a free particle or field quantum
\begin{eqnarray}
\exp\left\{-{\tau\over2p^+}
{({\bf q}_M-{\bf q}_0)^2}\right\}&=&
\int DcDbD{\bf q}\ e^{-S_0}\\
S_0&=&\int_0^\tau d\tau\int_0^{p^+} d\sigma
\left(b^\prime c^\prime -{1\over2}
{\bf q}^{\prime2}\right),
\label{lcevolve}
\end{eqnarray}
where the prime denotes $\partial/\partial\sigma$, and where
Dirichlet boundary conditions ${\bf q}(0,\tau)={\bf q}_0$,
${\bf q}(p^+,\tau)={\bf q}_M$, $b=c=0$ are imposed on
the world sheet fields.
The target space coordinates ${\bf q}(\sigma,\tau)$
are related to the transverse momentum carried by the system
by ${\bf p}=\int_0^{p^+}d\sigma {\bf q}^\prime={\bf q}_M-{\bf q_0}$,
and the Dirichlet boundary conditions on ${\bf q}$ ensure
the conservation of total momentum.
The ghosts $b,c$, which are Grassmann variables, are necessary
to ensure the correct measure factors. We shall always understand
path integrals to be the continuum limit of ordinary integrals
over variables defined on a lattice \cite{gilest}. 
We specify the lattice spacing in $\tau$ to
be $a$ and that in $\sigma$ to be $m$. The continuum limit
will be $a, m\to0$ with $m/a$ fixed. Then the measure of the
path integral is given by
\bea
DcDbD{\bf q}\equiv\prod_{j=1}^N
\prod_{i=1}^{M-1}{dc_i^jdb_i^j\over2\pi}
{d{\bf q}_i^j},
\eea
where $\tau=Na$ and $p^+=Mm$, with $M,N$ large positive integers.
Note that since $\sigma$
has dimensions of momentum and $\tau$ has dimensions of time,
$m/a$ has the dimensions of force. 

As discussed in \cite{bardakcit}, a general planar diagram
in quantum field theory can be represented as a path integral
similar to (\ref{lcevolve}) but with any number of internal
Dirichlet boundaries given by any number
of parallel line segments at fixed $\sigma$,
summed over varying location and length. Cubic vertices
are simply represented by the appearance or disappearance
of a Dirichlet boundary, and so are characterized locally on the
world sheet. Vertices of higher order are represented
by the simultaneous appearance and disappearance of
more than one Dirichlet boundary, and hence involve
nonlocal constraints on the geometry of the world sheet.

If we follow a line at fixed $\sigma$ in a general planar
diagram, we find a sequence of appearances and disappearances
of Dirichlet boundaries. Thus each such line is associated with a
two state system: the Dirichlet boundary is either ``on'' (solid line) or
``off'' (dotted line). Thus, in addition to the target space and ghost
worldsheet fields, we also introduce an Ising spin variable
$s_i^j=\pm1$ at each site of the worldsheet lattice. A time
link joining two $+$ spins is a bit of Dirichlet boundary, and
one joining two $-$ spins is a bit of bulk. A time link joining
opposite sign spins is a spin flip that turns a bit of
Dirichlet boundary on or off. The sum over all spin configurations
then accomplishes the sum over all planar diagrams.

In practice, the specification of the interacting Ising/target space
system involves case by case technical details, given for $\phi^3$ theory
in \cite{bardakcit} and for Yang-Mills in \cite{thornsheet}.
Here we only quote the final proposal for the $\phi^3$ theory
which is the focus of the rest of this article. The amplitude for
the sum of all planar diagrams evolving an initial state $i$
to a final state $f$ is given by\footnote{This formula is
Eq.~(26) of Ref.~\cite{thornsheet}, itself a refinement of 
the one given originally in \cite{bardakcit}. It is written
with a new, more transparent,
arrangement of the dependence on the lattice constants.}
\bea
T_{fi}&=&\lim_{\epsilon\to0}
\sum_{s_i^j=\pm1}\int DcDbD{\bf q}
\exp\left\{{{a\over 2m\epsilon}\sum_{i,j}b_{i}^{j}
c_{i}^{j}\sqrt{2m\epsilon\over a\pi}\left[{\cal V}_{0i}^j{\cal P}_i^j
+{\bar{\cal V}}_{0i}^j{\bar{\cal P}}_i^j
\phantom{\sum}\hskip-.5cm\right]}
\right\}
\nonumber\\&&
\exp\left\{-{a\over2m}\sum_{i,j}\left[({\bf q}_{i+1}^j-{\bf q}_{i}^{j})^2
+{({\bf q}_{i}^j-{\bf q}_{i}^{j-1})^2}
{P_i^jP_i^{j-1}\over\epsilon}\right]
+\sum_{i,j}
{a\over m\epsilon} P_i^{j-1}P_i^jP_i^{j+1}b^j_{i}c^j_{i}\right\}
\label{isingsumepsilon}\\
&&\exp\left\{{a\over m}\sum_{i,j}\left[(b_{i+1}^j-b_{i}^j)(c_{i+1}^j-c_{i}^j)
(1-P_i^j)(1-P_{i+1}^j)+(1-P_i^j)(P_{i+1}^j+P_{i-1}^j)b_i^jc_i^j
\right]\right\},\nonumber
\eea
where $P_i^j=(1+s_i^j)/2$ and we have put $\epsilon=m\epsilon/a$. 
The states $\ket{i},\ket{f}$ are specified by the
number and position of Dirichlet boundaries that extend
to $\tau=-\infty, +\infty$ respectively.
Here the vertex functions are given by
\bea
{\cal V}_{0i}^{j}&\equiv& {ga\over4m\sqrt{\pi}}
\exp\left\{-{a\over m}(b_{i-1}^{j}c_{i-1}^{j}
+b_{i+1}^{j}c_{i+1}^{j})\right\}\nonumber\\
{\bar{\cal V}}_{0i}^{j}&\equiv& {ga\over4m\sqrt{\pi}}\exp\left\{-{a\over m}
(b_{i+1}^{j+1}-b_{i}^{j+1})
(c_{i+1}^{j+1}-c_{i}^{j+1})\right\}.
\label{vertices}
\eea
Also the projectors multiplying the vertex functions are
\bea
{\cal P}_i^j\equiv P_i^jP_i^{j+1}(1-P_i^{j-1})\Pi_{i-1}^{j-1}\Pi_{i+1}^{j-1},
\qquad
{\bar{\cal P}}_i^j\equiv P_i^jP_i^{j-1}(1-P_i^{j+1})\Pi_{i-1}^{j}\Pi_{i+1}^{j},
\label{vertexprojectors}
\eea
where $\Pi_i^j\equiv(1+s_i^js_i^{j+1})/2$. 

Without going into great detail we make a few explanatory
comments about this formula. First, notice that the spin
projectors $P_i^j$ keep track of the distinction between
solid ($P=1$) and dotted ($P=0$) lines. A bit of Dirichlet 
boundary is represented by a delta function, identifying
the ${\bf q}$'s at successive sites, defined through the limit
\bea
\delta(\Delta{\bf q})=\lim_{\epsilon\to0}
\left({a\over2\pi m\epsilon}\right)^{D/2}
e^{-a\Delta{\bf q}^2/2m\epsilon}.
\eea
The prefactors in this formula are taken care of by the
ghost integration. Since formally $\delta({\bf 0})=V_\perp/(2\pi)^{D}$,
the volume of transverse space, we should
regard $\epsilon$ as a temporary infrared
cutoff on the transverse coordinates, related to the
size of the space $L=V^{1/D}$ by $\epsilon=2\pi a/mL^2$.   
The different arrangement of 
projectors for ghosts compared to those for ${\bf q}$ 
is due to the fact that
$b=c=0$ on solid lines, whereas the 
${\bf q}$'s are only required to be equal on
solid lines. The apparently elaborate ghost
dependence of the vertex functions in (\ref{vertices})
provides necessary factors of $1/p^+$ that are dictated
by the field theoretic Feynman rules. The noteworthy feature
is that in spite of their complexity, they are described
{\it locally} on the world sheet. The rather complicated combination
of projectors multiplying these vertex functions limits
their occurrence to the appearance or disappearance of 
Dirichlet boundaries. The extra $\Pi$ projectors remove
difficulties, due to the ghost insertions
in (\ref{vertices}), that occur when two or more 
vertices are within one or two lattice sites of each other.

Finally we comment on the limits involved in recovering
standard field theoretic
perturbation theory from Eq.~\ref{isingsumepsilon}.
One first expands the formula in a power series in $g$,
$T=\sum_n g^n T_n(\epsilon, a, m)$. Then the limit $\epsilon\to0$
on each $T_n$ converts it to a standard Feynman integral,
with uv cutoff $a$ and $p^+$ cutoff $m$. The dependence
on these cutoffs then just parameterizes the standard
field theoretic divergences. In this article we are attempting to 
learn about physics at finite coupling by studying
(\ref{isingsumepsilon}) in the presence of all these cutoffs.
But to compare our results to perturbation theory we must
keep in mind that the perturbation expansion is carried
out before sending $\epsilon\to0$, implying 
the parametric regime $g\ll\epsilon$.
\section{The Mean Field Approximation on a Continuous Worldsheet
with Simple Cutoff}
In this section, we will develop 
a somewhat crude, 
but qualitatively transparent, 
version of the mean field method on the world sheet. 
A more precise and detailed
version of the same method, based on the worldsheet lattice,
will be presented in the next section. Here we will adopt the
continuum version of the world sheet, and
we will use the notation and conventions of the previous section,
except that we use the
Minkowski metric in
both target space (so the factor at each vertex is $ig$, $g=$coupling const.)
and on the world sheet. The worldsheet coordinate $\sigma$ is compactified
in the interval 0 to $p^{+}$, and
 $\tau$ runs over the interval $\tau_{i}$
to $\tau_{f}$, which will eventually go to infinity. The target space
fields ${\bf q}(\sigma,\tau)$, which will be designated as matter
fields,  live in $D$ dimensions. For
simplicity, we take the total transverse momentum flowing through the
worldsheet to be zero:
$$
\int^{p^{+}}_{0} d\sigma {\bf q'(\sigma,\tau)}=
{\bf q}(p^{+},\tau)-{\bf q}(0,\tau)=0.
$$
To keep track of the two types of lines, it is convenient to define
a scalar
field $\phi$ on the worldsheet which is 1 on solid lines and 0 on dotted
lines. Now consider the expression
\be
\left(i\beta\phi\over\pi\right)^{D/2}\, \exp\left(-i\beta \phi
({\bf V})^{2}\right)
\label{1}
\ee
As $\beta$ tends to $\infty$, this expression tends to a delta function,
$\delta({\bf V})$, for all $\phi\neq 0$. Here,
we assume a Euclidean continuation to have a well defined Gaussian.
We will use this to impose the condition
$$
\dot{{\bf q}}=0
$$
on solid lines, where $\phi=1$. On dotted lines, where $\phi=0$,
there is no constraint. The matter part of the action, with the
above constraint, takes the form
\bq
S_{q}&=&\intst\, \mathcal{L}_{q},\nonumber\\
\mathcal{L}_{q}&=& \frac{1}{2}\beta\phi
 (\dot{{\bf q}})^{2}-\frac{1}{2}({\bf q}')^{2}.\label{2}
\eq
For the time being, we will keep the parameter
 $\beta$ finite, though eventually
we will let $\beta\rightarrow \infty$. Comparing to the lattice
expression (\ref{isingsumepsilon}), we see that $\phi$
here is playing the role of the projectors $P_i^jP_i^{j-1}$,
and $\beta=a^2/m^2\epsilon$.
Also, we have dropped
the prefactors in front of the exponential in (\ref{1}); these give rise
to quantum effects which will not play a role in the approximate
treatment of this section. However, their effect will be
included in the lattice calculations of Sec. 4, where we shall
see that they are mostly canceled by corresponding factors
in the ghost sector.

In addition to the matter part of the action, there is the
ghost action. The function of ghosts is to cancel the contribution
of the matter fields on the dotted lines, and leave it untouched
on solid lines. The  ghost action is somewhat
complicated, and its lattice version will be
given in the next section.  For our purposes,
we will not need the explicit form of the ghost action.
In fact, we find it easier to compute directly the effective
actions $W_{q,g}(\phi)$ for matter and ghost sectors as a
function of $\phi$. The calculation proceeds as follows:
The world sheet can be thought of as a union of black regions,
composed of only the solid lines, and the white regions, composed
of only the dotted lines (See Fig.1).
\begin{figure}[t]
\vskip1cm
\centerline{\epsfig{file=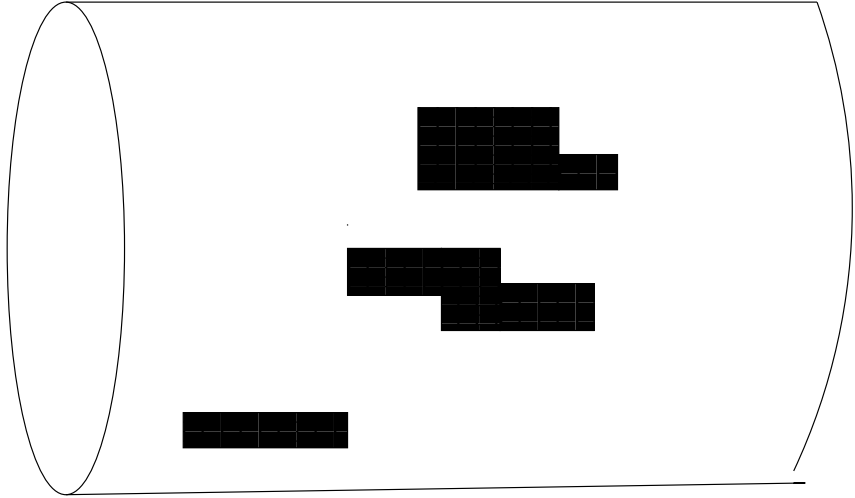,width=12cm}}
\caption{Regions of Worldsheet}
\vskip1cm
\end{figure}
 In the white regions, where
$\phi=0$, the situation is very simple: The matter and the ghost
contributions cancel, and the total effective action is zero:
\be
W_{q}(\phi=0)+W_{g}(\phi=0)=0.
\label{3}\ee
On the other hand, in the black regions, where $\phi=1$, the ghosts
do not contribute:
\be
W_{g}(\phi=1)=0.
\label{4}\ee
Therefore, we need only compute $W_{q}(\phi=1)$ for an arbitrarily
shaped black region. This is a standard Casimir type calculation.
Since the answer is divergent, a suitable cutoff has to be
introduced, and for the Casimir effect, one has to calculate
the cutoff independent finite part. Here, we have a much simpler
problem: We are interested in the dominant bulk contribution to
the action, which is going to be the leading cutoff dependent term.
It is also going to be proportional to the area of the black region
and independent of its shape.
We will compute this term
 for a rectangular black region of width
$L$ in the $\sigma$ direction and length $T$ in the $\tau$
direction, taking $\phi$ to be a constant ($\sigma$ and $\tau$
independent) in Eq.(\ref{2}). We will also
  impose periodic boundary conditions in both
directions and use a cutoff  in the mode number.
  Defining the operator $K$ by
$$
K= \beta\phi\partial_{\tau}^{2}-\partial_{\sigma}^{2},
$$
 we have
\be
W_{q}=-\frac{D}{2}\Tr \log(K).
\label{5}\ee
The cutoff dependent part of the action is easily calculated:
\be
W_{q}\cong -\frac{D\,L T}{4 \pi (\beta\phi)^{1/2}
\delta}.
\label{6}\ee
Here $\delta$, which has the dimensions of $\sigma^2$, is the cutoff 
parameter. 
It is proportional to $m^2$, where $m$ is
the discrete unit of $\sigma$ on the world sheet lattice.  
 
We note that
 the answer depends only on the area $L T$ of the region.
Although the calculation was done for a rectangular region, it
is clear that the leading cutoff dependent term is proportional to
the area even for a region of an arbitrary shape, and independent
of the boundary conditions imposed. This dependence on the area,
 and also the dependence on the combination $\beta\phi$,
 can be established essentially by dimensional and scaling
arguments, independent of any detailed calculation. The numerical
factor $4\pi$ is unimportant and it could be absorbed into
the definition of the cutoff.
 For small sized regions perimeter and 
shape dependent contributions may also become important.
 However, in this section, we will focus on the bulk
contribution only, and we will neglect corrections of this type.
The cutoff $\delta$ hides our ignorance of the texture of the
world sheet when the continuum limit is taken; it will eventually
be absorbed into the definition of the coupling constant.
Finally, the matter contribution to the action in the black
region is obtained by setting $\phi=1$ in Eq.(\ref{6}):
\be
W_{q}(\phi=1)=- \frac{D\,A}{4\pi (\beta)^{1/2} \delta},
\label{7}\ee
where $A$ is the area of the region. Eqs.(\ref{3}),(\ref{4})
 and (\ref{7}) provide
us with complete information about the contributions of the 
matter and ghost actions in the two types of regions.

Next, we have to tackle the functional integration over $\phi$.
We impose the constraint
that $\phi=0,1$
by means of a Lagrange multiplier
$\pi(\sigma,\tau)$:
\bq
\Delta S_{n}&=& \intst \,\Delta\mathcal{L}_{n},\nonumber\\
\Delta\mathcal{L}_{n}&=& \pi\left(\dot{\phi}- \sum_{1}^{n}
(-1)^{m}\delta(\tau-\tau_{m}(\sigma_{0}))\right).
\label{8}\eq
The functional integration over $\phi$ consists of  integration
over $\tau_{m}(\sigma_{0})$ and summation over n.
These summations and integrations
cannot be carried out in closed form; however,
the problem can be reduced to the solution of a differential equation.
Let us set
$$
\exp\left(\Delta S^{i,j}\right)= F^{i,j}(\tau_{i},\tau_{f}),
$$
where $\Delta S^{i,j}$ is the resulting action after the sums and integrals are
carried out. The outcome is one of the four possible functions on the
right hand side, depending on the boundary conditions on $\phi$ at $\tau_{i}$
and $\tau_{f}$. The indices $i$ and $j$ can take on the values 
($+$) or ($-$), 
corresponding to $\phi=1,0$ at $\tau=\tau_{i,f}$. These functions are
defined through the differential equations
\bq
\partial_{\tau}F^{+,+}(\tau_{i},\tau)&=& ig\: \exp\left(i\pi(\tau)\right)
\,F^{+,-}(\tau_{i},\tau),\nonumber\\
\partial_{\tau}F^{+,-}(\tau_{i},\tau)&=& ig\: \exp\left(-i\pi(\tau)\right)
\,F^{+,+}(\tau_{i},\tau),\nonumber\\
\partial_{\tau}F^{-,+}(\tau_{i},\tau)&=& ig\: \exp\left(i\pi(\tau)\right)
\,F^{-,-}(\tau_{i},\tau),\nonumber\\
\partial_{\tau}F^{-,-}(\tau_{i},\tau)&=& ig\: \exp\left(-i\pi(\tau)\right)
\,F^{-,+}(\tau_{i},\tau),
\label{9}\eq
plus the initial conditions
\bq
F^{+,+}(\tau_{i}=\tau_{f})&=& F^{-,-}(\tau_{i}=\tau_{f})=0,\nonumber\\
F^{+,-}(\tau_{i}=\tau_{f})&=& F^{-,+}(\tau_{i}=\tau_{f})=1.
\label{10}\eq
In writing these equations, we have assumed that the vertex that
converts a dotted line into a solid line or vice versa is simply
given by the coupling constant $g$. It was shown in
 \cite{bardakcit} that there are
additional contributions to the vertex involving the ghosts $b$ and $c$;
however, in the leading large $D$ limit, which will be the
approximation scheme adopted in this article,
they do not contribute. This is because the function of these
vertex ghost insertions is to
produce the factor of $1/p^{+}$, which is part of the propagator, and it
does this by deleting a pair of ghost fields $b$ and $c$, as explained in 
\cite{bardakcit}.
The crucial point is that it is always a pair of fields irrespective
of the transverse dimension $D$, since $p^{+}$ always appears to the 
first power and not to the power $D$. Therefore, this contribution
goes like 1 and not $D$, and it is negligible in the large $D$ limit.
We will have more to say about the large $D$ limit later.
 
It turns out to be convenient, although not essential, to keep
track of the $(+)$ and ($-$) indices by means of two fermionic
(anticommuting) variables $e_{1,2}(\tau,\sigma)$. These fermions,
 already introduced in \cite{bardakcit}, give the continuum formulation of
the Ising system of the next section.
 $e_{1}$ is associated with the $(+)$ index, or the solid lines,
and $e_{2}$ with the ($-$) index, or the dotted lines. Their $\tau$
dependence is given by
\bq
e_{1}(\tau,\sigma)&=&F^{+,+}(\tau_{i},\tau) e_{2}(\tau_{i},\sigma)
+ F^{-,+}(\tau_{i},\tau) e_{1}(\tau_{i},\sigma),\nonumber\\
e_{2}(\tau,\sigma)&=& F^{+,-}(\tau_{i},\tau) e_{2}(\tau_{i},\sigma)
+ F^{-,-}(\tau_{i},\tau) e_{1}(\tau_{i},\sigma).
\label{11}\eq
As a result, the $e$'s satisfy the differential equations
\bq
i\dot{e}_{1}+ g\; \exp(i\pi(\tau)) e_{2}&=&0,
\nonumber\\
i\dot{e}_{2}+ g\; \exp(-i\pi(\tau)) e_{1}&=&0,
\label{12}\eq
leading to the action
\be
S_{f}=\intst\left(i\bar{e}_{1}\dot{e}_{1}
+i \bar{e}_{2}\dot{e}_{2}+ g\; \exp(i\pi)
\bar{e}_{1} e_{2}+ g\; \exp(-i\pi) \bar{e}_{2}
e_{1}\right).
\label{13}\ee
For later application, it is convenient to make the change of variables,
\bq
e_{1}&=& \exp\left(\frac{i}{2}\pi\right)\psi_{1},\; \bar{e}_{1}=
\exp\left(-\frac{i}{2}\pi\right)\bar{\psi}_{1},\nonumber\\
e_{2}&=&\exp\left(-\frac{i}{2}\pi\right)\psi_{2},\; \bar{e}_{2}=
\exp\left(\frac{i}{2}\pi\right)\bar{\psi}_{2},
\label{14}\eq
with the corresponding action
\be
S_{f}=\intst\left(i\bar{\psi}_{1}\dot{\psi}_{1}+ i\bar{\psi}_{2}
\dot{\psi}_{2}-\frac{1}{2}\dot{\pi}\,\bar{\psi}_{1}\psi_{1}
+\frac{1}{2}\dot{\pi}\,\bar{\psi}_{2}\psi_{2} + g\, \bar{\psi}_{1}
\psi_{2}+ g\,\bar{\psi}_{2}\psi_{1}\right).
\label{15}\ee
Adding all the contributions, the full expression for 
$$
T_{fi}=\langle {\bf p}_{f}|e^{-\tau p^{-}}|{\bf p_{i}}\rangle
$$
is
\bq
T_{fi}&=&\int D{\bf q}\,D\lambda\,D\phi\,Db\,Dc\nonumber\\
&&\times \exp\left(i\intst(\mathcal{L}_{q}+\mathcal{L}_{g}
+\mathcal{L}_{f}-\dot{\pi}\phi)\right),
\label{16}
\eq
where  $\mathcal{L}_{q}$ and $\mathcal{L}_{f}$ are given by Eqs.(\ref{2})
and (\ref{15}).

We are interested in the ground state of the system described by the
action given above. We will compute the ground state of the
combined matter, ghost and the fermionic
systems in the presence of constant fixed
background fields $\phi$ and $\pi$, and then minimize the total
energy with respect to these background fields by solving the
classical equations of motion.
 This is in essence the mean field method.
 A systematic way to do this is to consider the large
$D$ limit, where $D$ is the number of transverse dimensions. In practice,
this number is not particularly large, so the method could at best
be expected to give qualitative results. However, it is a convenient
way of organizing a systematic expansion scheme in inverse powers
of $D$. We note that the contributions of the matter and ghost fields
 to the ground state energy are proportional to $D$, so 
it is convenient to scale the field $\pi$ and the coupling
constant $g$ by
\be
\pi\rightarrow D\pi,\;g\rightarrow D g,
\label{17}\ee
in order to have an effective action proportional to $D$. In what
follows, we will simplify things by considering only the leading
contribution in $D$. This does not mean that the non-leading
contributions are unimportant; for example, it is clear that
 Lorentz invariance can only be understood by taking into
account the non-leading terms. Also, there is the question of 
the meaning of the coupling constant $g$. From Eq.(\ref{9}), one sees
that this constant always has the dimensions of  mass
(or energy). The dimension of the constant $g$ that appears in
the $g\,\phi^{3}$ interaction in field theory is ${\rm mass}^{(4-D)/2}$,
depending on the dimensionality of space-time; 
only for $D=2$ (4 dimensional space-time), does
it have the dimensions of mass. As a result, we cannot identify
the constant $g$ that we are using with the field theory
coupling constant, except perhaps in four dimensional 
space-time. In other dimensions, we have to treat it as an
effective parameter, and to establish its connection with
the parameters in field theory would again require a knowledge
of higher order terms in the $1/D$ expansion
On the light-cone worldsheet $\sigma$ has dimensions of
momentum and $\tau$ the dimensions of time. On our
worldsheet lattice, the
lattice constants in the two directions therefore have a
dimensionful  ratio $m/a$, with dimensions of force. Since
this ratio may be kept fixed in the limit of a
continuous world sheet, the formalism provides a scale which
can be used to define a dimensionless coupling constant.
${\hat g}\equiv g\sqrt{1/32\pi^2}(a/m)^{(4-D)/4}$. The coupling constant
used in (\ref{9}) can therefore be identified with
${\hat g}(m/a)^{(4-D)/4}$ up to dimensionless factors.

Another important simplification follows from the nature of the
ground state. Since
the problem is translationally invariant in both the compactified
$\sigma$ and the $\tau$ direction, at least in the limit
$\tau_{f}-\tau_{i}\rightarrow \infty$, we expect the ground state
to share these symmetries; therefore,
 we take $\phi_{0}$, the
expectation value of $\phi$, to be independent of $\sigma$ and
$\tau$. We note that
 the  value of the product $\beta\phi_{0}$ is especially
significant: From Eq.(\ref{2}), it follows that a finite
non-zero value of this product leads to the standard string
action with a finite slope parameter. The key idea behind our
computation is self consistency. Starting with a finite
$\beta\phi_{0}$, we compute the zero point energy of the
resulting string, and add this to the energy of the fermions
 to get the total energy. Minimizing this energy,
we arrive at a finite value of $\beta\phi_{0}$, completing
the cycle of self consistency.

In trying to compute the energy of the combined matter and 
ghost system for a background value $\phi_{0}$ of $\phi$,
we encounter a problem. We know the effective action for this system,
from which the ground state energy is easily deduced, 
only
for $\phi=0$ and $\phi=1$ (Eqs.(\ref{3}),(\ref{4}), and (\ref{7})). 
On the other hand, $\phi_{0}$ can take on any value between 0 and 1, so
we have to extend the definition of the action to an arbitrary
$\phi_{0}$ between these limits. This can be done as follows:
$\phi_{0}$ is the classical expectation value, or the average
value of $\phi$. Consider a specific partitioning of the world
sheet between the white and black regions, such as represented
by figure 1. If we denote the total area of the black regions
by $A_{b}$ and the total area of the world sheet by $A_{w}$,
and remembering that $\phi=0$ on the white regions and
$\phi=1$ on the dark regions,
the average value of $\phi$, $\phi_{0}$, for this partitioning
is given by
\be
\phi_{0}=\frac{A_{b}}{A_{w}}.
\label{18}
\ee
On the other hand, from Eqs.(\ref{3}),(\ref{4})
 and (\ref{7}), the contribution
of this partitioning to the combined matter and ghost action is
\be
W_{q}+W_{g}=- \frac{D\,A_{b}}{4\pi(\beta)^{1/2}\delta}.
\label{19}\ee
From these equations, it follows that
\be
W_{q}+W_{g}=-\frac{DA_{w}\,\phi_{0}}{4\pi(\beta)^{1/2}\delta},
\label{20}\ee
which is the equation that expresses the combined action in terms
of $\phi_{0}$. It is now easy to convert this into an equation
for the corresponding ground state energies, which we label as
$E_{q}^{(0)}$ and $E_{g}^{(0)}$ respectively. Remembering that
the area of the world sheet is given by
$$
A_{w}= p^{+} (\tau_{f}-\tau_{i}),
$$
we have,
\be
E_{q}^{(0)}+E_{g}^{(0)}=\frac{D\,p^{+}\phi_{0}}{4\pi
(\beta)^{1/2}\delta}.
\label{21}\ee
The combined matter and ghost energy therefore has a simple
linear dependence on $\phi_{0}$. This is a direct consequence
of the area law for the matter action (see Eq.(\ref{7})). Since this
simple dependence on area is bound to get corrections for
small regions, we expect to have some deviation from the
linear dependence of the energy on $\phi_{0}$. In fact, a
calculation based on the worldsheet lattice, presented in the
next section, results in a more complicated dependence, but
the linear dependence can be regarded as a reasonable approximation.

Next, we have to introduce some background for the field $\pi$.
At first, one might think that, by translation invariance, this should
again be a constant. However, it turns out that a constant background
is trivial; it is clear from Eq.(\ref{15}) that
 only the derivative of $\pi$ with respect to $\tau$
has dynamical significance.
 So we ansatz 
\be
\langle \pi\rangle\cong \pi^{(0)}(\tau)=
\pi_{0}\tau+\pi_{1},
\label{22}\ee
where $\pi_{0,1}$ are constants. Again, we see from Eq.(\ref{15}) that
this background, which at first looks time dependent, is in fact 
static, and the system has a well defined energy.
 To compute this
energy, it is convenient to quantize the fermion action of Eq.(\ref{15}),
in the background given by Eq.(\ref{22}), and construct the Hamiltonian.
 The result is
\bq
H_{f}&=&\int_{0}^{p^{+}} d\sigma\left(\frac{1}{2}\pi_{0}
\bar{\psi}_{2} \psi_{2} -\frac{1}{2}\pi_{0} \bar{\psi}_{1}
\psi_{1} -g\bar{\psi}_{1}\psi_{2} - g\bar{\psi}_{2}\psi_{1}
\right)\nonumber\\
&\rightarrow&\sum_{n}\left(\frac{1}{2}\pi_{0}\, a_{n,2}^{\dagger}
a_{n,2} -\frac{1}{2}\pi_{0}\, a_{n,1}^{\dagger}a_{n,1} - g\,
a_{n,1}^{\dagger} a_{n,2} -g\, a_{n,2}^{\dagger} a_{n,1}\right).
\label{23}\eq
In this equation, $a_{n,i}$ and $a_{n,i}^{\dagger}$ are fermionic
operators that satisfy the usual anticommutation relations
$$
\{a_{m,i},a_{n,j}^{\dagger}\}_{+}=\delta_{i,j}\delta_{m,n},
$$
where $i,j=1,2$. The fermionic field has the mode expansion
\bq
\psi_{i}&=&\sum_{n}(p^{+})^{-1/2}\, a_{n,i}\,
 \exp\left(\frac{2\pi i n\sigma}
{p^{+}}\right),\nonumber\\
\bar{\psi}_{i}&=&\sum_{n} (p^{+})^{-1/2}\, a_{n,i}^{\dagger}\,
\exp\left(-\frac{2\pi i n\sigma}{p^{+}}\right).
\label{24}\eq
The vacuum is as usual annihilated by the $a$'s. In the state
under consideration, each mode number n is filled by one of the
$a_{n,i}^{\dagger}$'s. This corresponds, according to the value of
$i$, to either having a solid or a dotted line. There is no meaning
to an unoccupied state.

The Hamiltonian of Eq.(\ref{23}) is easily diagonalized. There are two 
eigenvalues for each mode n:
\be
\kappa_{n}^{\pm}= \pm\frac{1}{2}(\pi_{0}^{2}+4 g^{2})^{1/2}.
\label{25}\ee
To find the energies corresponding to the original fermionic
variables $e_{1,2}$, we have to perform the transformation
of Eq.(\ref{14}), which results in four possible energies for each
mode:
\be
\epsilon^{(n)}_{f,\pm,\pm}=\pm\frac{1}{2}\pi_{0}
\pm\frac{1}{2}(\pi_{0}^{2}+4 g^{2})^{1/2}.
\label{26}\ee
These four different energies correspond to four different
boundary conditions which can be imposed on the fermionic
state at the initial time. However, we shall see that  only
$(-,+)$ and $(-,-)$ combinations satisfy the constraint
$$
0\leq \phi\leq 1.
$$
We will consider both of these possibilities in what follows.

 The next step is the computation
of the total fermionic energy, which, as it stands, will be
divergent. One way to regularize it is to discretize the
$\sigma$ interval, with a spacing $\Delta\sigma$ between two
adjacent points. It is then easy to see that, to go from
the constraints of Eq.(\ref{8}) written in discrete form, to an
integral over $\sigma$, we need a factor of $\Delta\sigma$,
which can be generated by a suitable scaling of $\pi_{0}$
and $g$:
$$
\pi_{0}\rightarrow \Delta\sigma\,\pi_{0},\;\;
g\rightarrow \Delta\sigma\;g.
$$
Alternatively, one can use a cutoff $\nu$ on the mode
number, the relation between the two
cutoffs being
$$
\Delta\sigma\rightarrow \pi\nu.
$$
Using this latter cutoff, the total fermionic energy is given by
\be
E_{f}=\frac{\Delta\sigma}{2}\sum_{n}\left(\pi_{0}-(\pi_{0}^{2}
+4 g^{2})^{1/2}\right) \exp\left(-\frac{2\pi |n|\nu}
{p^{+}}\right)\rightarrow \frac{p^{+}}{2}\left(\pi_{0}+
(\pi_{0}^{2}+4 g^{2})^{1/2}\right),
\label{27}\ee
where, at the end, we have taken the limit of small $\nu$.

\begin{figure}
\begin{center}
\psfrag{'4g^21.01'}{{\tiny ${ 4g^2=0.01}$}}
\psfrag{'4g^2.1'}{{\tiny ${ 4g^2=0.1}$}}
\psfrag{'4g^21'}{{\tiny ${ 4g^2=1.0}$}}
\psfrag{'pi_0'}{{ $\pi_0$}}
\psfrag{'phi_0'}{{ $\phi_0$}}
\epsfig{file=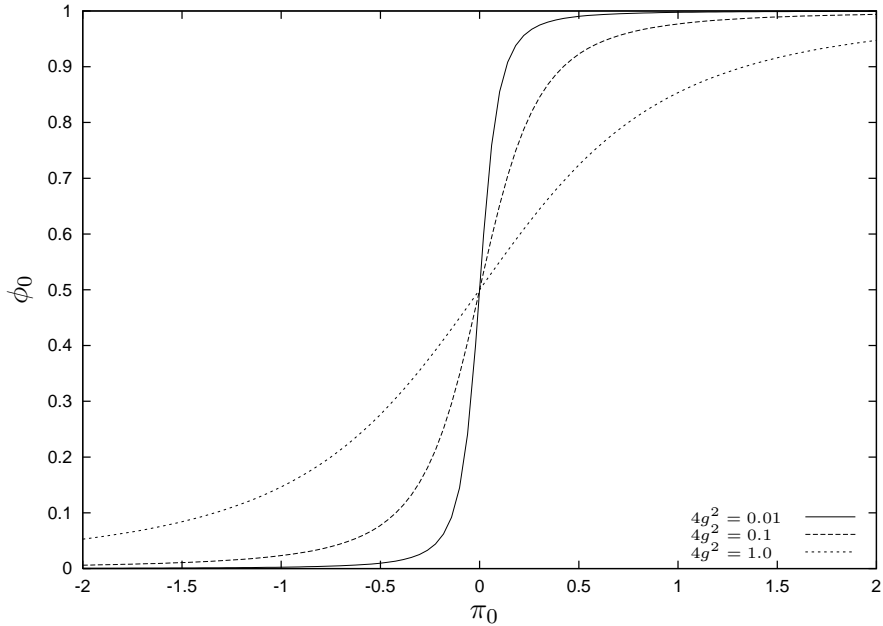,width=12cm}
\caption{The mean field $\phi_0$ as a function $\pi_0$, 
as given by Eq.~(\ref{29}) with the choice of $-$ sign,
for various couplings.}
\label{phikofpik}
\end{center}
\end{figure}

The total ground state energy is then the sum of $E_{q}^{(0)}$, $E_{g}^{(0)}$,
$E_{f}$ and the contribution from the last term in the
exponential in Eq.(\ref{16}):
\be
E_{0}^{\pm}=D\,p^{+}\left(\frac{\phi_{0}}{4\pi
(\beta)^{1/2}\delta}
+\pi_{0}\phi_{0}-\frac{1}{2}\pi_{0}\pm
\frac{1}{2}(\pi_{0}^{2}+4 g^{2})^{1/2}\right).
\label{28}\ee
To find the ground state energy, we have to minimize $E_{0}^{\pm}$
with respect to $\pi_{0}$ and $\phi_{0}$. Minimizing it with
respect to $\pi_{0}$ gives
\be
\frac{1}{D p^{+}}\frac{\partial E_{0}^{\pm}}{\partial\pi_{0}}
=-\frac{1}{2}\pm \frac{\pi_{0}}{2(\pi_{0}^{2}
+4 g^{2})^{1/2}}+ \phi_{0}=0.
\label{29}\ee
We note that, as $\pi_{0}$ ranges between $-\infty$ and
$+\infty$, the range of $\phi_{0}$ is between 0 and 1, as it should be.
We plot $\phi_0(\pi_0)$ in Fig.~\ref{phikofpik}.
Had we chosen the $(+,+)$ or $(+,-)$ combinations in Eq.(\ref{26}), we
would not have the correct range for $\phi_{0}$.
Next, minimizing the energy with respect to $\phi_{0}$ gives
\be
\frac{1}{D p^{+}}\frac{\partial E_{0}^{\pm}}{\partial \phi_{0}}
=\frac{1}{4\pi(\beta)^{1/2}\delta}
+\pi_{0}=0.
\label{30}\ee

We will first consider these equations in the weak coupling
limit and compare the results with perturbation theory.
Accordingly, we will assume $g$ to be much smaller than $\pi_{0}$,
and expand Eq.(\ref{28}) in lowest order in their ratio:
\bq
\phi_{0}&\cong&\frac{1}{2}\pm \left(\frac{1}{2}-\frac{g^{2}}
{\pi_{0}^{2}}\right)\nonumber\\
&=&\frac{1}{2}\pm\left(\frac{1}{2}- 16\pi^{2} g^{2}
\beta\delta^{2}\right).
\label{31}\eq
Since in the weak coupling limit we expect the dotted lines
(white regions) to dominate, $\phi_{0}$ should be small
compared to one, and the minus sign is the correct choice
in the above equation:
\be
\phi_{0}(weak)\cong 16\pi^{2} g_{0}^{2}\beta,
\label{32}\ee
where we have scaled the coupling constant according to
$$
g_{0}=g\delta.
$$
This corresponds to the combination $(-,-)$ in Eq.(\ref{26}), and from
Eq.(\ref{28}), we see that it has lower energy compared to the alternative
combination $(-,+)$. It is also clear from Eq.(\ref{31}) that $g_{0}$
must satisfy the condition
\be
4\pi g_{0} (\beta)^{1/2}\ll 1
\label{33}\ee
for $\phi_{0}$ to be small, and also, as was assumed, for
$g$ to be smaller than $\lambda_{0}$.
 For the sake of completeness,
we note that the combination $(-,+)$, which has higher energy
and is therefore unstable, corresponds to $\phi_{0}\approx 1$
and a resulting predominance of solid lines.

We do not expect perturbative field theory to lead to string
formation, so it is important to verify this
 in the weak coupling regime. As we have remarked
earlier, strings form only if the parameter $\phi_{0}\beta$
is non-zero and finite as $\beta\rightarrow\infty$.
 If we define the weak coupling limit by the
condition
$$
g_{0}\beta\rightarrow 0,
$$
then, from Eq.(\ref{32}), we see that
 there is no string formation. In fact, it follows from Eq.(\ref{2})
that, as expected,
 this is the zero slope or the field theory limit.

The next question is if or when string formation is realized.
This happens, at least in the mean field approximation, in
a regime which we will call the intermediate coupling regime.
$g_{0}$ still satisfies the condition (\ref{32}), but now if we 
require that
$$
g_{0}\beta\neq 0,
$$
which means
$$
\beta\phi_{0}\neq 0.
$$
This is the condition for a finite non-zero slope,
and quantization of $S_{q}$ then results in the
usual string spectrum with the level spacing given by
$\pi/((\beta\phi_{0})^{1/2} p^{+})$. In passing,
 we note that the presence of
$p^{+}$ in the denominator is consistent with Lorentz invariance.
Remembering that we are calculating the spectrum of $p^{-}$ with
${\bf p}$ set equal to zero, the product $p^{+} p^{-}$ has to
be a Lorentz invariant pure number, as is the case here. Of
course, all of our results are conditional on the validity
of the mean field approximation.

Finally, we would like to discuss briefly the strong coupling regime,
by which we mean the range of values of $g_{0}$ that do not
satisfy the condition given by (\ref{33}). For example,
 $g_{0}$ could remain
finite and non-zero as $\beta\rightarrow\infty$. This corresponds
to the infinite slope limit of the string: The level splitting goes
to zero, and the whole string spectrum collapses into a continuum.

 The last topic we wish to discuss is the ratio of the density of
dotted lines to the density of the solid lines.
 To compute this ratio, which we call r, we need the ground
state of $H_{f}$. It is given by
\be
|s\rangle= \prod_{n}\left(u_{n}\, a_{n,1}^{\dagger} + v_{n}\,
a_{n,2}^{\dagger}\right)|0\rangle,
\label{34}\ee
where the constants $u_{n}$ and $v_{n}$ satisfy
\be
\frac{v_{n}}{u_{n}}=- \frac{1}{2 g}\left((\pi_{0}^{2} +
4 g^{2})^{1/2} +\pi_{0}\right),\;\;\; |u_{n}|^{2}+
|v_{n}|^{2}=1.
\label{35}\ee
The ratio r for state is then given by
\be
r=\frac{|\langle s|\int_{0}^{p^{+}} d\sigma\,\bar{\psi}_{2}
\psi_{2} |s\rangle |}{|\langle s|\int_{0}^{p^{+}} d\sigma\,
\bar{\psi}_{1}\psi_{1}|s\rangle |}
=\frac{|v_{n}|^{2}}{|u_{n}|^{2}}
=\frac{1}{4 g^{2}}
\left((\pi_{0}^{2}+ 4 g^{2})^{1/2} -\pi_{0}\right)^{2}
=\frac{1-\phi_{0}}{\phi_{0}}.
\label{36}\ee
Since $\pi_{0}$ is negative (Eq.(\ref{30})), this ratio is always
greater than one, so there are always more dotted lines than
solid lines. In the weak and intermediate coupling regimes, we
can approximately write
\be
r\approx \frac{1}{16\pi^{2} g_{0}^{2}\beta}.
\label{37}\ee
In the weak coupling limit, $g_{0}\beta\rightarrow 0$, so
$r\rightarrow\infty$, even for a finite $\beta$.
 As expected, this means that the worldsheet
is taken over by the dotted lines, and there are hardly any
solid lines.  In the intermediate coupling
regime, $g_{0}^{2}\beta$ is finite and non-zero, so r is
finite for finite $\beta$, although as $\beta\rightarrow\infty$,
it is still true that $r\rightarrow\infty$. So the dotted lines
are still in preponderance, although not to the same extent as
in the case of weak coupling. It is somewhat surprising that,
even in the presence of this relative scarcity of solid lines,
string formation can take place.

\section{Mean Field Path Integral on the Worldsheet Lattice}

We would now like to apply the mean field method
directly to the lattice system (\ref{isingsumepsilon}).
It is this system which we have shown is equivalent to the
summation of planar diagrams \cite{bardakcit}. In
particular the worldsheet lattice spacings $a,m$ are
directly related to cutoffs in $x^+$ and $p^+$ that
were employed to render all Feynman integrals finite.
 
In applying mean field theory to Eq.~\ref{isingsumepsilon}, 
we make the same simplification as in
Section 3 by dropping the
extra ghost insertions at the ends of solid lines (\ie by
setting the exponential factors in (\ref{vertices}) to
unity).  As explained in Section 3, mean field theory is expected to
be valid when $D\to\infty$ where $D$ is the number
of $q$ fields or twice the number of ghost $b,c$ pairs.
In such a limit, this simplification
should have negligible effect at $D=\infty$,
since the dropped ghost insertions involve only one of the 
infinite number of ghost pairs. Once this simplification is
made, the extra $\Pi$ projectors in (\ref{vertexprojectors}) can
be safely removed, and we can then make the replacements
\bea
{\cal V}_{0i}^j{\cal P}_i^j
+{\bar{\cal V}}_{0i}^j{\bar{\cal P}}_i^j\to
{ga\over4m\sqrt{\pi}}P_i^j[P_i^{j+1}+P_i^{j-1}-2P_i^{j-1}P_i^{j+1}].
\eea
Because of the projectors, the interaction terms appear only when the
ghosts multiplying them are decoupled from all other ghosts. This means
we can remove the factors 
\bea
{ga\over4m\sqrt{\pi}}\sqrt{2m\epsilon\over a\pi}
=g\sqrt{\epsilon}
\sqrt{a\over8m\pi^2}\equiv 2\sqrt{\epsilon}{\hat g}
\equiv {\hat g}^\prime
\eea
from the exponent and then regain the correct answer by inserting the
factor $\prod_{ij}({\hat g}^\prime)^{(1-s_i^js_i^{j-1})/2}$ in
the path integral. Here we have also defined the dimensionless
couplings ${\hat g}$ and ${\hat g}^\prime$.

\noindent These manipulations lead to the simplified expression
\bea
T^{\rm simp}_{fi}&=&\lim_{\epsilon\to0}
\sum_{s_i^j=\pm1}\int DcDbD{\bf q}
\prod_{i,j}\left(2{\hat g}\sqrt{\epsilon}\right)^{(1-s_i^js_i^{j-1})/2}
\nonumber\\&&\hskip-.25cm
\exp\left\{-{a\over2m}\sum_{i,j}\left[({\bf q}_{i+1}^j-{\bf q}_{i}^{j})^2
+{({\bf q}_{i}^j-{\bf q}_{i}^{j-1})^2}
{P_i^jP_i^{j-1}\over\epsilon}\right]
+\sum_{i,j}
{a\over 2m\epsilon}P_i^j(P_i^{j+1}+P_i^{j-1})b^j_{i}c^j_{i}\right\}
\nonumber\\
&&\hskip-.25cm
\exp\left\{{a\over m}\sum_{i,j}\left[(b_{i+1}^j-b_{i}^j)(c_{i+1}^j-c_{i}^j)
(1-P_i^j)(1-P_{i+1}^j)+(1-P_i^j)(P_{i+1}^j+P_{i-1}^j)b_i^jc_i^j
\right]\right\}
\label{simpsum}
\eea
Note that the factor $\sqrt\epsilon$ multiplying
${\hat g}$ is a direct consequence of the mismatch of
Dirichlet enforcing delta functions on each solid
line segment: the value of ${\bf q}$ on the solid line
is integrated, whereas the ghosts are set to zero everywhere
on the solid line. Therefore, the prefactors in the 
Gaussian approximation to the delta functions do not
quite cancel. 

To facilitate a mean field treatment we introduce unity in the
form 
\bea
1=\int D\phi\prod_{ij}\delta(\phi_{ij}
-P_i^j)=\int D\phi D\lambda \exp\left\{i\sum_{ij}\lambda_{ij}(\phi_{ij}
-P_i^j)\right\}
\eea
into the functional integrand, and then assume we can treat the
$\lambda$ and $\phi$ integrals classically. That is, in this
section we identify
the mean field approximation with a saddle point evaluation of
the integrals over $\phi$ and $\lambda$. One advantage of
this way of implementing the approximation is that it sets
the stage for a systematic exploration of fluctuation effects
that certainly correct, and in some cases (as for example with the
two dimensional Ising model) 
invalidate some of the results of the mean field approximation.

The classical treatment of the $\lambda$ integration introduces
some freedom in what we take as the zeroth order in the 
approximation scheme. It is this integration that constrains
the values of $\phi_{ij}$ to be 0 or 1, and approximating the 
$\lambda$ integration will relax these constraints to some degree.
Thus valid rearrangements of the defining path integral (\ref{simpsum}),
which exploit the fact that the $P$'s are projectors, will
lead to different classical (\ie\ zeroth order) actions.
The fundamental premise of the mean field approximation is
that slowly varying, and in particular uniform, 
field configurations capture the
important physics of the system. If this is really true, 
the results should be insensitive to reasonable rearrangements
along these lines. If there is sensitivity, comparing the results of 
different starting points can give some measure of
the credibility of the approximation scheme. The most naive
setup is to blindly substitute $\phi_{ij}$ for each
$P_i^j$ with no further rearrangement. 

How can we decide on the best zeroth order action? Of course, 
we would like to choose the one
which gives the closest possible agreement
with the actual answer. For coupling the mean field
to the matter and ghosts, we can get some guidance from
a simple exact evaluation of Eq.~\ref{isingsumepsilon}
at zero coupling. Then all solid lines are eternal,
and we can easily calculate the
exact energy for $n$ equally spaced solid lines that
all extend from early to late times. For $n/M$ fixed
this should correspond to the energy at uniform
mean field $\phi=n/M$. By construction, the formula
gives for the integral over all variables on dotted lines
\bea
\prod_{j=1}^N\left({a\over2\pi m\epsilon}\right)^{nD/2}\exp\left\{
-{a\over2m}\sum_{i=1}^n
{({\bf q}_{i+1}^j-{\bf q}_{i}^j)^2\over(M/n)}
-{a\over2m\epsilon}\sum_{i=1}^n({\bf q}_{i}^j-{\bf q}_{i}^{j-1})^2\right\},
\eea
where these ${\bf q}$'s are those on the solid
lines. Integrating over all of them gives
a lattice string path integral of the type exactly
evaluated in \cite{gilest} and briefly discussed in the
appendix. The answer for the
bulk energy is
\bea
aE_n\equiv M{\cal E}_{q,g}=D\sum_{l=1}^n\sinh^{-1}\left(\sqrt{\epsilon n\over M}\sin{l\pi\over n}
\right)&\to& 
n{2D\over\pi}{\rm Re}\left\{i{\rm Li}_2\left({1\over i}
\sqrt{n\epsilon\over M}\right)\right\}\nonumber\\
&=&M{2D\over\pi}\phi
{\rm Re}\left\{i{\rm Li}_2\left({-i}\sqrt{\phi\epsilon}\right)\right\}
\label{exact0}\\
&\sim&M{2D\over\pi}\sqrt{\epsilon}\phi^{3/2}\label{smalleps},
\eea
where in the first line we have taken $M$ large, in the second line
identified $n=M\phi$, and in the last line taken $\epsilon$ small.
There are of course other configurations of eternal solid lines
that could simulate a uniform mean field, but this one
has the lowest energy.

As far as mean field calculations go, given their
relative crudeness, taking the
exact zero coupling result (\ref{exact0}) as the matter and ghost 
contribution to the energy, or even taking the linear
interpolation used in Section 3, is probably as reliable
as other choices.
However, if we want to use the saddle point
formulation to systematically go beyond this approximation,
it is useful to have a zeroth order action that
leads to a qualitatively similar answer. Also, we
would like to identify the mean field in some way with an
effective string tension, and for this we need to know
at least how the target space fields couple to the mean field
in the path integral action. Finally, once interactions come
into play, the lowest zero coupling energy given by
(\ref{exact0}) need no longer be the appropriate energy to
assign to the ghost and matter system, and other possibilities
should be kept in mind.

\begin{figure}
\begin{center}
\psfrag{'phi'}{$\phi$}
\psfrag{'cale'}[bl][tr][1][0]{${\cal E}_{qg}$}
\psfrag{'calendm'}{\small naive}
\psfrag{'calend'}{\small naive, $\mu=0$}
\epsfig{file=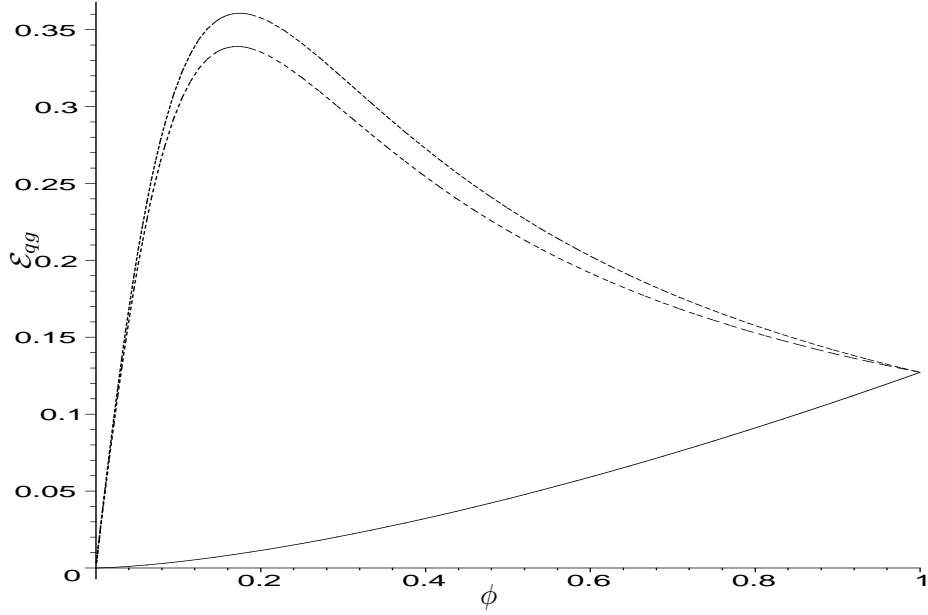,width=12cm,height=8cm}
\caption{Comparison of the ghost$+$matter energy for the
naive zeroth order action (dashed lines) 
to the zero coupling result of Eq.~\ref{exact0} (solid line)
at $\epsilon=0.01$. The higher dashed curve has  the ghost
mass parameter $\mu$ set to zero.}
\label{comparenaive}
\end{center}
\end{figure}

Even so, the result (\ref{exact0})
is strong support for immediately disposing of the naive choice 
for zeroth order action, which leads to the bulk ghost-matter energy density
\bea
{\cal E}^{\rm naive}_{q,g}={2D\over\pi}{\rm Re}
\left\{i{\rm Li}_2\left({\sqrt{\epsilon}\over i\phi}\right)\right\}
-{D}\ln\left(\sqrt{{1\over4}+\mu}
+\sqrt{{1\over4}+\mu
+{\epsilon(1-\phi)^2\over\phi^2}}\right),
\label{enaive}
\eea
where $\mu={\epsilon(1-\phi)(2-\phi)/2}$. In Fig~\ref{comparenaive},
we plot on the same graph this result, the same
expression with $\mu=0$, and the exact zero coupling one.
We see that the curves for (\ref{exact0}) and
(\ref{enaive}) disagree in
almost every respect qualitatively and
quantitatively. Clearly, we should find a better
starting point! 

One problem with the naive starting point is that the
non-gradient ghost terms do not have a transparent
formal continuum limit. This bodes ill for
a slowly varying mean field approximation.
We can improve this situation with two
rearrangements. First consider the change of
ghost variables $b_i^j,c_i^j\to b_i^j-P_i^jb_i^{j-1},c_i^j- P_i^jc_i^{j-1}$,
which has unit Jacobian. 
Then we see that the only change in Eq.~\ref{simpsum} is 
the replacement of the last term in the second line,
\bea
{a\over 2m\epsilon}P_i^j(P_i^{j+1}+P_i^{j-1})b^j_{i}c^j_{i}
\to{a\over 2m\epsilon}P_i^j(P_i^{j+1}+P_i^{j-1})(b^j_{i}-b_i^{j-1})
(c^j_{i}-c_i^{j-1}).
\label{dotghosts}
\eea
The shift has no effect on the other ghost terms because they
contain at least one factor of $1-P_i^j$. Now the
formal continuum limit of (\ref{dotghosts})
is transparent. As stressed in
Section 3, the function of the ghosts is to cancel the
effects of the matter fields on dotted lines (where $\phi_{ij}=0$). 
This change makes the cancelation more effective
for small but nonzero mean field as well. We shall refer
to the ghosts treated this way as {\it dynamical ghosts},
and shall restrict the calculations in this section to that
case.  Secondly, we can rearrange the last term
of the last exponent in (\ref{simpsum}) which as it stands also has
an obscure formal continuum limit. Note the identity
\bea
(1-P_i^j)(P_{i+1}^j+P_{i-1}^j)={1\over2}\left[P_{i+1}^j+P_{i-1}^j
-2P_i^j +(P_i^j-P_{i+1}^j)^2+(P_i^j-P_{i-1}^j)^2\right].
\eea
Although it takes more space to write, the right side is formally of
order $m^2$, so the formal continuum limit is
apparent. In addition this rearrangement causes this
``ghost mass term'' to vanish for uniform
mean fields. In all that follows, we shall adopt both of
these rearrangements. 

If we make only the rearrangements mentioned thus far, substitution
of $\phi_{ij}=\phi$ leads to a ghost and matter path integral
action in which $a\phi^2/m\epsilon$ multiplies the time difference terms
of both matter and ghosts, and $a/m$ ($a(1-\phi)^2/m$) multiply
the space difference terms of the matter (ghosts). The upshot of 
doing the path integrals with this action is the bulk energy
\bea
{\cal E}^1_{q,g}={2D\over\pi}{\rm Re}
\left\{i{\rm Li}_2\left({\sqrt{\epsilon}\over i\phi}\right)\right\}
-{2D\over\pi}{\rm Re}
\left\{i{\rm Li}_2\left({\sqrt{\epsilon}(1-\phi)
\over i\phi}\right)\right\}.
\eea  
Here we see the basic delicacy with the saddle point approach
to the mean field approximation. The result agrees with the
exact answer at $\phi=1$, but the $\phi\to0$ behavior depends on
a delicate cancelation between the matter and ghost contributions.
In particular, unlike the exact calculation, the limits
$\epsilon\to0$ and $\phi\to0$ do not commute.

\begin{figure}
\begin{center}
\psfrag{'phi'}{$\phi$}
\psfrag{'cale'}[l][r][1][0]{${\cal E}_{qg}$}
\psfrag{'cale1'}{\tiny${\cal E}_{qg}^1$}
\psfrag{'cale2'}{\tiny${\cal E}_{qg}^2$}
\psfrag{'cale3'}{\tiny${\cal E}_{qg}^3$}
\epsfig{file=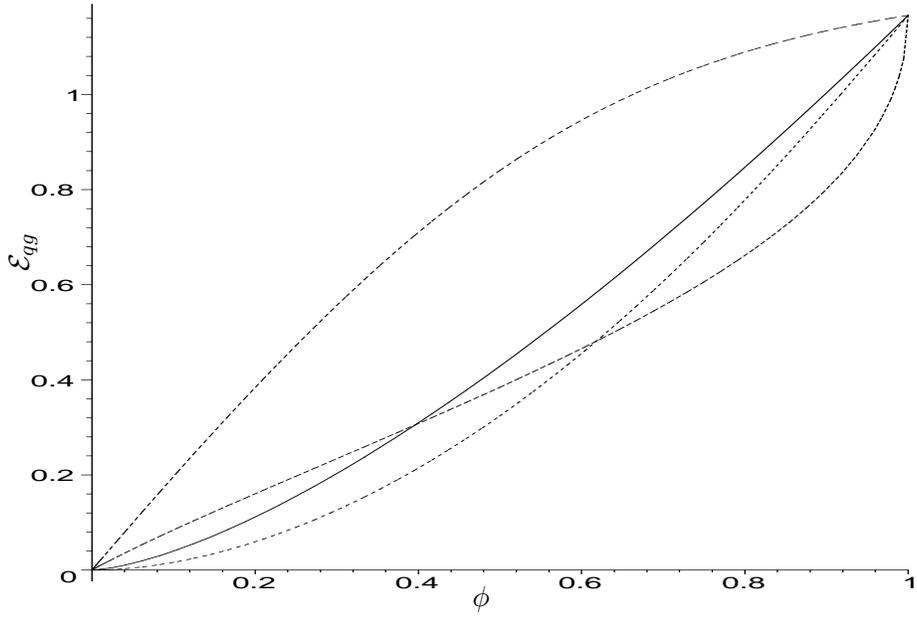,width=12cm,height=8cm}
\caption{Comparison of ghost$+$matter energies for $\epsilon=1.0$. The solid
line is the exact zero coupling energy (\ref{exact0}), the highest dashed
line is ${\cal E}^1_{qg}$, the dashed line with infinite slope at
$\phi=1$ is ${\cal E}^2_{qg}$, and the third dashed line is the compromise
${\cal E}^3_{qg}$.}
\label{compgh1}
\end{center}
\end{figure}

\begin{figure}
\begin{center}
\psfrag{'phi'}{$\phi$}
\psfrag{'cale'}[l][r][1][0]{${\cal E}_{qg}$}
\epsfig{file=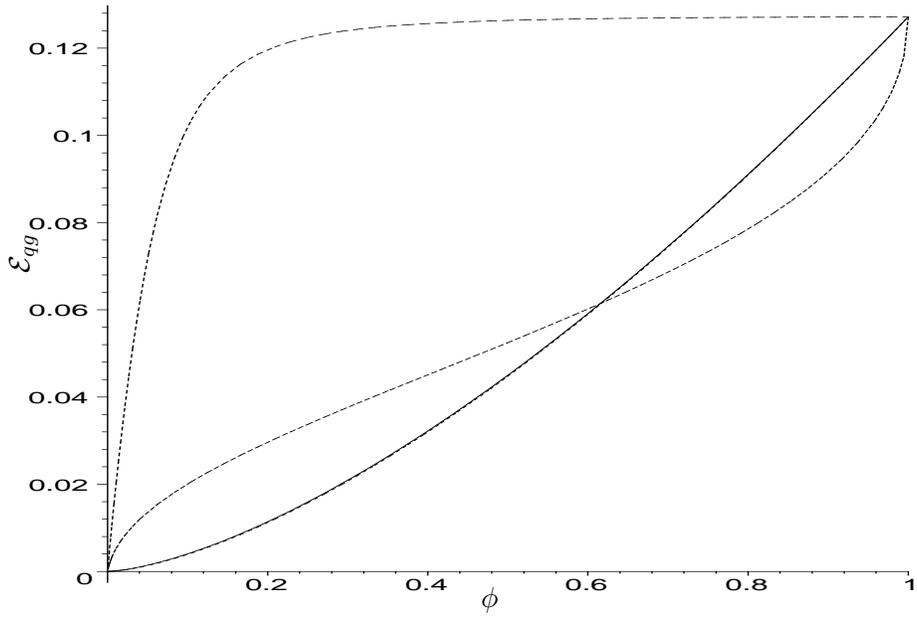,width=12cm,height=8cm}
\caption{Comparison of ghost$+$matter energies at $\epsilon=0.01$.}
\label{compgh2}
\end{center}
\end{figure}

Another reasonable choice for zeroth order action emerges from
rearranging the products of projectors with the identity
\bea
P_i^{j}P_i^{j-1}={P_i^{j}+P_i^{j-1}\over2}-{(P_i^{j}-P_i^{j-1})^2\over2}
\to{\phi_i^{j}+\phi_i^{j-1}\over2}-{(\phi_i^{j}-\phi_i^{j-1})^2\over2},
\label{lineardot}
\eea
and we could similarly write
\bea
(1-P_{i+1}^{j})(1-P_{i}^{j})=1-{P_{i+1}^{j}+P_i^{j}\over2}
-{(P_{i+1}^{j}-P_i^{j})^2\over2}.
\eea
If we do both these things we arrive, for uniform mean fields, 
at the bulk energy
\bea
{\cal E}^2_{q,g}={2D\over\pi}{\rm Re}
\left\{i{\rm Li}_2\left({1\over i}{\sqrt{\epsilon\over\phi}}\right)\right\}
-{2D\over\pi}{\rm Re}
\left\{i{\rm Li}_2\left({1\over i}{\sqrt{\epsilon(1-\phi)
\over \phi}}\right)\right\}.
\eea
We compare the different possibilities in Fig~\ref{compgh1}
for $\epsilon=1.0$ and in Fig.~ \ref{compgh2} for $\epsilon=0.01$.

We see that in Fig.~\ref{compgh2} the curve for a compromise
form ${\cal E}^3$ 
\bea
{\cal E}^3_{q,g}\equiv{2D\over\pi}{\rm Re}
\left\{i{\rm Li}_2\left({\sqrt{\epsilon}\over i\sqrt{\phi}}\right)\right\}
-{2D\over\pi}{\rm Re}
\left\{i{\rm Li}_2\left({(1-\phi^2)\sqrt{\epsilon}
\over i\sqrt{\phi}}\right)\right\}.
\eea
lies virtually on top of the exact answer. This is because the squaring
of $\phi$ in the second term
makes both curves approach zero like $\phi^{3/2}$ and
the squaring of $(1-\phi^2)$ removes the 
square root branch point at $\phi=1$. The zeroth order action
that leads to this interpolation is given by the
substitution $(1-P_i^j)(1-P_{i+1}^j)\to(1-\phi_i^{j2})(1-\phi_{i+1}^{j2})$
in front of the spatial difference term for ghosts, and the substitution
(\ref{lineardot}) in front of the time difference terms for
both matter and ghosts. It thus involves a valid but
non-minimal interpretation of the projectors
appearing here. Because it does well with
the zero coupling case, we shall 
take this zeroth order action in the mean field calculations to follow.

Recall that the saddle-point approach
to the mean field approximation consists of replacing integration over
$\phi_{ij}, \lambda_{ij}$ with the classical equations of motion
for these variables. This procedure factorizes the
remaining quantum averages into the Ising sum times the
Gaussian integrals over ${\bf q},b,c$. To study the ground state
we look for a solution with constant $\phi_{ij}=\phi$ and
$\lambda_{ij}=\lambda$. The formal continuum limit of the
the zeroth order matter and ghost action described in the 
previous paragraph is for uniform\footnote{The formal
continuum limit for slowly varying non-uniform fields
has only one new term:
\bea
S_{q,g}= \int d\tau d\sigma \left({1\over2}{\bf q}^{\prime2}
+{a^2\phi\over 2m^2\epsilon}{\dot{\bf q}}^2-(1-\phi^2)^2b^\prime c^\prime
-{a^2\phi\over m^2\epsilon}{\dot b}{\dot c}
-{1\over2}(\phi^{\prime\prime}+2\phi^{\prime2})bc\right).\nonumber
\eea} mean field $\phi$
\bea
S_{q,g}= \int d\tau d\sigma \left({1\over2}{\bf q}^{\prime2}
+{a^2\phi\over 2m^2\epsilon}{\dot{\bf q}}^2-(1-\phi^2)^2b^\prime c^\prime
-{a^2\phi\over m^2\epsilon}{\dot b}{\dot c}
\right)
\label{contaction}
\eea
from which we see that the effective string tension as a function
of the mean field is $T_{eff}(\phi)=m\sqrt{\epsilon}/a\sqrt{\phi}$.
In the following we also use $\beta\equiv a^2/m^2\epsilon$,
in terms of which the effective tension reads 
$T_{eff}(\phi)=(\sqrt{\beta\phi})^{-1}$. We are of course
interested in the limit $\beta\to\infty$, which, if taken at
fixed $\phi$, implies $T_{eff}\to0$. However, the 
perturbative field theory is characterized by a worldsheet
which in the bulk has no $\dot{\bf q}^2$ term which means
$T_{eff}\to\infty$. The perturbative situation will be recovered
if we find that, as a function of $\beta$, $\phi$ tends to 0
faster than $1/{\beta}$.

As a final preliminary to the mean field calculation, we note that
the Ising sum can also be done exactly for constant mean field
$\lambda$. The sums for each $i$ factorize, so we have
\bea
\sum_{s_i^j=\pm1}
\left({\hat g}^\prime\right)^{(1-s_i^js_i^{j-1})/2}
e^{{-i\sum_{ij}\lambda P_i^j}}=\left(\sum_{s^j=\pm1}
\left({\hat g}^\prime\right)^{(1-s^js^{j-1})/2}
e^{{-i\sum_{j}\lambda P^j}}\right)^{M}.
\eea
where we put ${\hat g}^\prime\equiv 2{\hat g}\sqrt{\epsilon}$
The sums over spins just amount to matrix multiplication of
the transfer matrix
\bea
{\cal T}\equiv\pmatrix{e^{-i\lambda}&{\hat g}^\prime e^{-i\lambda}\cr
{\hat g}^\prime&1\cr}
\eea
by itself $N$ times. The ground state energy is thus $-M/a$ times the
$\ln$ of the largest eigenvalue of this matrix.
The eigenvalues are easily found
\bea
t_{\pm}={1+e^{-i\lambda}\pm\sqrt{(1-e^{-i\lambda})^2
+4{\hat g}^{\prime2} e^{-i\lambda}}\over2},
\eea
with the corresponding unnormalized eigenvectors given by
\bea
v_{\pm}=\pmatrix{1\cr 
(t_\pm e^{i\lambda}-1)/{\hat g}^\prime\cr}
\eea
We shall find that the classical equations for $\phi,\lambda$
imply that $i\lambda$ is real, in  which case $t_+$ is the
largest eigenvalue. We therefore find that the ground
state energy of the Ising system is
\bea
aE_s=-M\ln{1+e^{\kappa}+\sqrt{(1-e^{\kappa})^2
+4{\hat g}^{\prime2} e^{\kappa}}\over2},
\eea
where we have put $-i\lambda\equiv\kappa$ in anticipation that
it is real. It is also of interest to give the ground
state expectation of $(1\pm s)/2$ given by 
\bea
\bra{G}{1+s\over2}\ket{G}&\equiv&
\lim_{N\to\infty}{v^{\prime T}{\cal T}^{N}\pmatrix{1&0\cr0&0}{\cal T}^{N}v
\over v^{\prime T}{\cal T}^{2N}v}={e^{-i\lambda}-t_-\over t_+-t_-}\\
\bra{G}{1-s\over2}\ket{G}&\equiv&
\lim_{N\to\infty}{v^{\prime T}{\cal T}^{N}\pmatrix{0&0\cr0&1}{\cal T}^{N}v
\over v^{\prime T}{\cal T}^{2N}v}={1-t_-\over t_+-t_-}\eea 
These two quantities give the mean number of solid lines and
dotted lines respectively.

Putting all of the contributions to the ground state energy
of the system together (with ${\cal E}^3_{qg}$ chosen
for the matter and ghost contribution), we have 
\bea
{\cal E}(\phi,\kappa)&\equiv&
{aE_{\rm total}\over M}=
{2D\over\pi}{\rm Re}
\left\{i{\rm Li}_2\left({\sqrt{\epsilon}\over i\sqrt{\phi}}\right)\right\}
-{2D\over\pi}{\rm Re}
\left\{i{\rm Li}_2\left({\sqrt{\epsilon}(1-\phi^2)
\over i\sqrt{\phi}}\right)\right\}\nonumber\\
&&+\kappa\phi-\ln{1+e^{\kappa}+\sqrt{(1-e^{\kappa})^2
+4{\hat g}^{\prime2} e^{\kappa}}\over2}.
\label{caledyn}
\eea

\begin{figure}
\begin{center}
\psfrag{'phi_0'}{$\phi$}
\psfrag{'pi_0'}{$\kappa$}
\epsfig{file=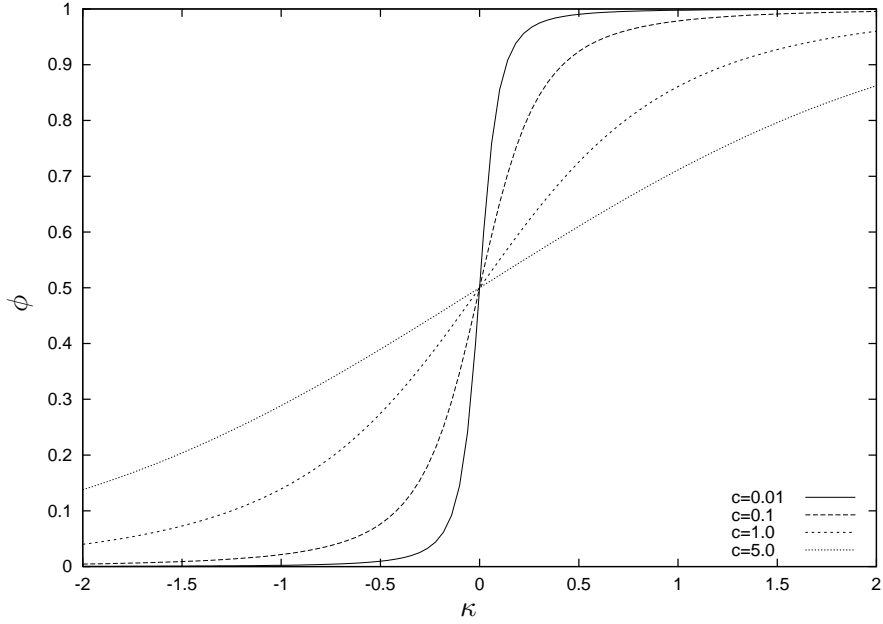,width=12cm}
\caption{The mean field $\phi(\kappa)$ of Eq.~(\ref{kappa0}) for
various couplings $c\equiv16{\hat g}^2\epsilon$. Note the similarity to
Fig.~\ref{phikofpik}.}
\label{phiofkappa}
\end{center}
\end{figure}

The last step is to minimize with respect to $\kappa,\phi$ and
take $\beta\to\infty$. We first minimize with respect to $\kappa$
obtaining 
\bea
\phi(\kappa)&=&{e^{\kappa}\over\sqrt{(1-e^{\kappa})^2
+4{\hat g}^{\prime2} e^{\kappa}}}
{-1+e^{\kappa}+2{\hat g}^{\prime2}
+\sqrt{(1-e^{\kappa})^2+4{\hat g}^{\prime2}e^{\kappa}}
\over 1+e^{\kappa}+\sqrt{(1-e^{\kappa})^2
+4{\hat g}^{\prime2}e^{\kappa}}}\\
&=&{e^{\kappa}-t_-\over t_+-t_-}
= {1\over2}+{\sinh(\kappa/2)\over2\sqrt{\sinh^2(\kappa/2)
+{\hat g}^{\prime2}}}
\label{kappa0}
\eea
which confirms that $\phi=\langle P_i^j\rangle$.
For ${\hat g}^{\prime2}\ll\sinh^2(\kappa/2)$, $\phi\to0$ or $1$ according
to whether $\kappa<0$ or $>0$. Thus $\phi(\kappa)$ 
is a smoothed approximation to $\theta(\kappa)$,
with the step getting sharper as ${\hat g}^\prime\to0$.
In Fig.~\ref{phiofkappa} we plot $\phi$ as a function of $\kappa$ 
for various values of ${c}=4{\hat g^{\prime2}}$.

To give an energetic interpretation of the solutions to these
equations, we first solve (\ref{kappa0}) for $\kappa$ calling
the solution $\kappa_0(\phi)$:
\bea
\kappa_0(\phi)=-2\sinh^{-1}\left({{\hat g}^\prime(1-2\phi)\over2
\sqrt{\phi(1-\phi)}}\right)
=-2\sinh^{-1}\left({{\hat g}\sqrt{\epsilon}(1-2\phi)\over
\sqrt{\phi(1-\phi)}}\right).
\eea
Then we substitute $\kappa_0(\phi)$ for $\kappa$ in the
energy to get ${\cal E}(\phi,\kappa_0(\phi))$ as a function of
$\phi$. In fact, the $\kappa$ dependent part of the
energy becomes simply
\bea
&&\kappa\phi-\ln{1+e^{\kappa}+\sqrt{(1-e^{\kappa})^2
+4{\hat g}^{\prime2} e^{\kappa}}\over2}=\nonumber\\
&&\hskip3cm \phi\kappa_0(\phi)
-\ln\left({\sqrt{\phi(1-\phi)+{{\hat g}^2\epsilon(1-2\phi)^2}}
+{{\hat g}\sqrt{\epsilon}}\over
 {\sqrt{\phi(1-\phi)+{{\hat g}^2\epsilon(1-2\phi)^2}}
+{{\hat g}\sqrt{\epsilon}}(1-2\phi)}}\right).
\eea

\begin{figure}
\begin{center}
\psfrag{'kappa'}{$-\kappa$}
\psfrag{'phi'}{$\phi$}
\psfrag{'b=0.01'}{\tiny$\epsilon=0.01\quad$}
\psfrag{'b=0.1'}{\tiny$\epsilon=0.1\quad\quad$}
\psfrag{'b=0.001'}{\tiny$\epsilon=0.001$}
\epsfig{file=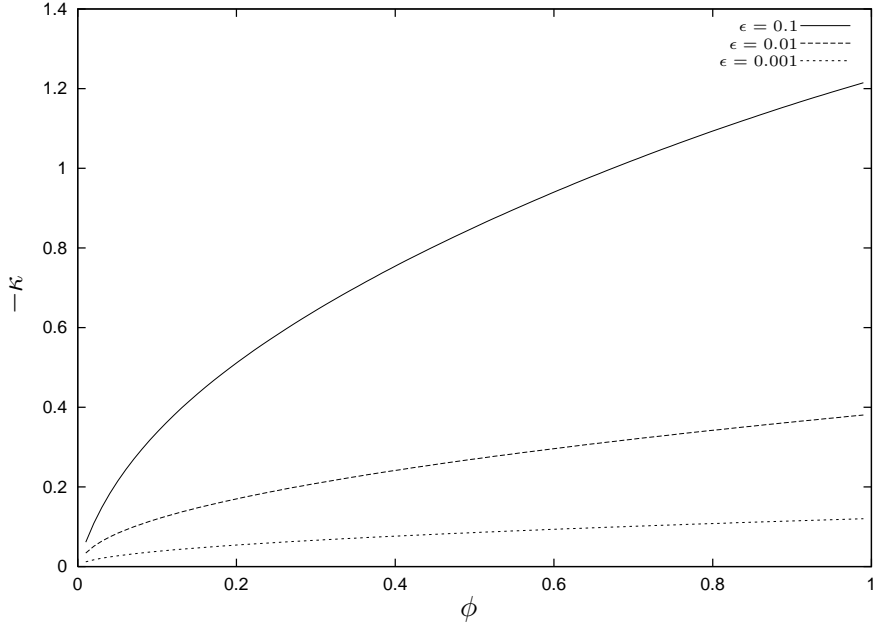,width=12cm}
\caption{The function $-\kappa(\phi)$ determined from the
minimization of the energy with respect to $\phi$ for various
values of $\epsilon$. We have set $D=2$ for this plot.}
\label{kappadofphi}
\end{center}
\end{figure}

\begin{figure}
\begin{center}
\psfrag{'cale'}[Bl][Bl][1][90]{${\cal E}$}
\psfrag{'phi'}{$\phi$}
\epsfig{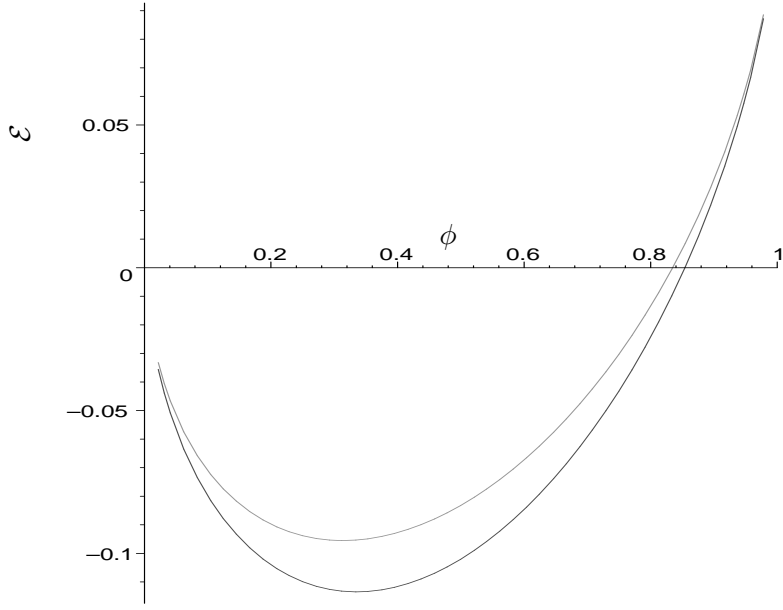}
\caption{The energy curve ${\cal E}(\phi,\kappa_0(\phi))$ for $\epsilon=0.01$
and $c=16{\hat g}^2\epsilon=0.1$ is the lower one. The upper curve
uses the linear interpolation for the
matter$+$ghost energy as in Section 3. We see that the differences
between the two treatments are quite mild. We have set $D=2$ for this plot.}
\label{caled101}
\end{center}
\end{figure}

Minimization with respect to $\phi$ gives a formula for $\kappa$
\bea
\kappa(\phi)={D\over\pi\phi}\left[\arctan
\left({\sqrt{\epsilon}\over \sqrt{\phi}}\right)-{1+3\phi^2\over1-\phi^2}
\arctan\left(
{\sqrt{\epsilon}(1-\phi^2)\over \sqrt{\phi}}\right)\right].
\eea
Clearly the stationary points of the energy as a function of
$\phi$ are given by
\bea
{d\over d\phi}{\cal E}(\phi,\kappa_0(\phi))=\kappa_0(\phi)-\kappa(\phi)=0.
\eea
In Fig.~\ref{kappadofphi} we plot $-\kappa$ for several values of $\epsilon$.
(in this and all the plots that follow we set $D=2$, corresponding
to 4 dimensional space-time.)
A noteworthy feature here is that $-\kappa$
stays positive over the whole range of $\phi$. As a consequence,
the solutions of $\kappa=\kappa_0$ will all be in the interval
$0\leq\phi\leq1/2$. In Fig.~\ref{caled101} we plot ${\cal E}(\phi,\kappa_0(\phi))$ for $b=0.01$ and $c=0.1$. It shows a single stationary point which
is a minimum.

\begin{figure}
\begin{center}
\psfrag{'eprime'}{${\cal E}^\prime/\sqrt{\epsilon}$}
\psfrag{'phi'}{$\phi$}
\psfrag{'0.01'}{\tiny${\hat g}=0.01$}
\psfrag{'0.1'}{\tiny${\hat g}=0.1$}
\psfrag{'1.0'}{\tiny${\hat g}=1.0$}
\epsfig{file=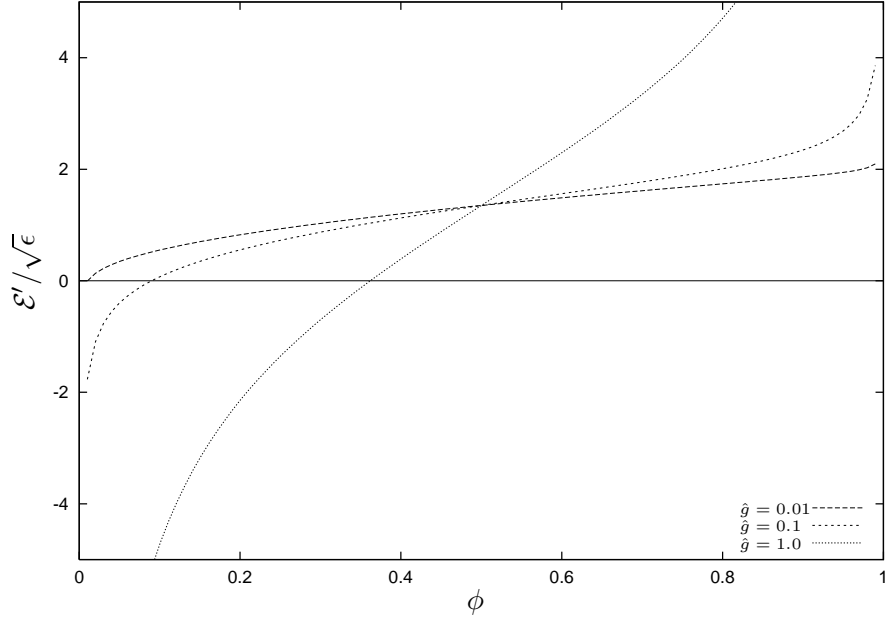,width=12cm}
\caption{The left side of Eq.~(\ref{fixedphieq}), which determines
the mean field at finite ${\hat g}$ for various couplings. Note that
all solutions are in the range $0\leq\phi\leq1/2$, with strong
coupling corresponding to the upper end of this range. We have set 
$D=2$ in this plot.}
\label{eprimedfixedphi}
\end{center}
\end{figure}

We can analyze the solutions analytically in the limit 
$\epsilon\to0$. Solutions with $\phi\neq0,1$ in the
limit are obtained by expanding $\kappa$ and $\kappa_0$:
\bea
\kappa\approx -{3D\sqrt{\epsilon\phi}\over\pi},
\qquad \kappa_0\approx-{2{\hat g}\sqrt{\epsilon}(1-2\phi)\over
\sqrt{\phi(1-\phi)}}.
\eea
The $\sqrt{\epsilon}$ cancels, and the equation for the minimum becomes
\bea
{3D\sqrt{\phi}\over\pi}-{2{\hat g}(1-2\phi)\over\sqrt{\phi(1-\phi)}}=0.
\label{fixedphieq}
\eea
The left side is plotted for various ${\hat g}$ values 
in Fig.~\ref{eprimedfixedphi}. As already mentioned, the solutions
all lie in the range $0\leq\phi\leq1/2$.
Notice that Eq.~(\ref{fixedphieq}) simplifies to a cubic equation
for $\phi$ with coefficients linear in ${\hat g}^2$. Thus the
solution $\phi({\hat g})$ as an analytic function of ${\hat g}$
will have branch points in the finite ${\hat g}^2$ complex plane,
and so its Taylor expansion about ${\hat g}=0$
will have a finite radius of convergence as expected for
the sum of planar diagrams \cite{thooftfiniteradius}.
By virtue of the identification of the string tension
$T_{eff}(\phi)=1/\sqrt{\beta\phi}
=m\sqrt{\epsilon}/(a\sqrt{\phi})$, we see that all solutions
for ${\hat g}$ finite as $\epsilon\to0$ correspond to
zero string tension. In other words the system falls apart
into a tensionless soup.

\begin{figure}
\begin{center}
\psfrag{'E/ep2'}{${\cal E}/{\epsilon}^2$}
\psfrag{'eta'}{$\eta$}
\epsfig{file=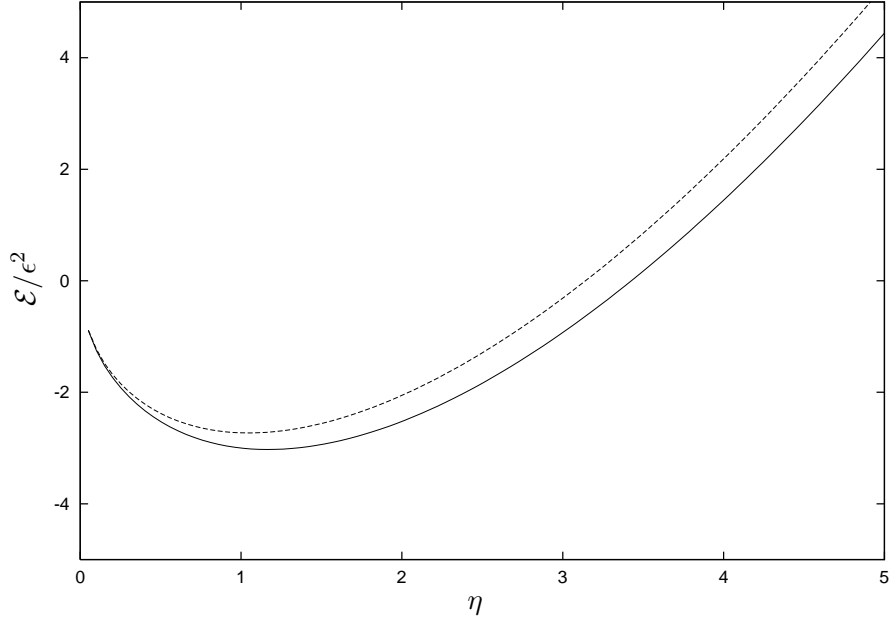,width=12cm}
\caption{Energy at fixed $\eta$ for $G={\hat g}/\epsilon=1$. 
The solid (lower) curve is from (\ref{energyfixedeta}) and
the dashed (higher) curve uses (\ref{exact0}) for the
matter$+$ghost energy. For this plot we set $D=2$.} 
\label{enfixedeta}
\end{center}
\end{figure}

Next we analyze the case for $\phi$ near zero. We define
a rescaled mean field by $\eta\equiv\phi/{\epsilon}$
and take $\epsilon\to0$ at fixed $\eta$. Note that this is
the combination that enters the effective
string tension $T_{eff}=1/\sqrt{\beta\phi}=m/\sqrt{a^2\eta}$, 
so a solution
with finite $\eta$ in the limit $\epsilon\to0$
would indicate a finite non-zero effective string tension
suggesting string formation. Then we find for the energy in
this limit
\bea
{\cal E}(\eta{\epsilon},\kappa)&\approx&
\epsilon\left(\epsilon{2D\over\pi}
\eta^2\arctan{1\over\sqrt{\eta}}
-2\eta\sinh^{-1}{{\hat g}\over\sqrt{\eta}}
-2\eta{{\hat g}\over{\hat g}
+\sqrt{\eta+{\hat g}^2}}\right).
\label{energyfixedeta}
\eea
Applying the same limit to $\kappa$, we find
\bea
\kappa(\epsilon\eta)
\approx -{D\epsilon\over\pi}
\left(4\eta\arctan{1\over\sqrt{\eta}}-{\eta^{3/2}\over1+\eta}\right),
\eea
and the slope of the energy 
is given by
\bea
{d{\cal E}(\phi,\kappa_0(\phi))\over d\phi}=-\kappa(\phi)+\kappa_0(\phi).
\eea
The stationary points are of course the solutions of $\kappa=\kappa_0$.
Since $\kappa=O(\epsilon)$, we can only get a solution at
fixed $\eta$ if we also have ${\hat g}=O(\epsilon)$.
Putting ${\hat g}=G\epsilon$, the stationarity condition reads
\bea
{D\over\pi}
\left(4\eta\arctan{1\over\sqrt{\eta}}-{\eta^{3/2}\over1+\eta}\right)-
{2G\over\sqrt{\eta}}=0.
\eea
It is clear from the fact that for any $G>0$, the left side goes
to $-\infty$ ($+\infty$) when $\eta$ goes to 0 ($\infty$), that there is
always a solution to this equation. Note incidentally that had
we used (\ref{exact0}) for the ghost$+$matter energy the first
term on the left would have been replaced by $3D\sqrt{\eta}/\pi$,
that is, by its asymptotic form at large $\eta$, and the
same qualitative conclusion applies. But there are quantitative
differences in the results.
For comparison of the energy derived from (\ref{exact0}) in the
fixed $\eta$ regime with ${\hat g}=O(\epsilon)$ to (\ref{energyfixedeta})
in the same regime we plot both energy curves for $G=1$ in
Fig.~\ref{enfixedeta}.

Finally we note that for $G\ll 1$ or ${\hat g}\ll\epsilon$, there is
a solution for $\eta\ll1$ given by $\eta\approx(G/D)^{2/3}$, which
corresponds to effective tension $T_{eff}\approx m(D/G)^{1/3}/a$.
This limit corresponds to the regime of perturbation theory
where the worldsheet fields in the bulk of the world sheet
are constrained in terms of the fields on the boundaries.

\section{Conclusions}
In this article, we have presented a scheme for 
approximately summing Feynman
graphs of arbitrary order. The scheme is an adaptation of the mean
field method, widely used in many body physics and field theory, to the
worldsheet representation of perturbation theory that we have developed
earlier. For simplicity, we chose $\phi^{3}$ as a toy model for study,
although we hope to apply the methods developed here to more realistic
models, such as non-abelian gauge theories. The goal was to investigate
under what circumstances string formation is possible in field 
theory. The worldsheet reformulation of field theory seems ideally
suited to such an investigation; in particular, the field $\phi$,
which naturally appears in this formulation and which roughly
 represents the density of propagating particles, is closely
related to the string tension. A non-zero expectation value for this
field signals formation of a string with finite tension.

The problem is  to find a reliable way to compute this expectation
value. In this article, we have used the mean field method as a
possible approach to this problem. In this approach, the fields on
the worldsheet are assigned various background expectation values,
which are then calculated self consistently by minimizing the 
total energy of the system. In this fashion, one can find out
whether the energetics can support a non-zero string tension.
Unfortunately, there are many technical difficulties in carrying
out this program. The biggest problem is connected with the
delicate cancelation between the matter and ghost sectors, where two
big numbers can cancel each other to give zero. Because of this,
the results of any approximate treatment can be somewhat unstable.
We try to bypass some of these problems by adopting a simple
minded approach in the continuum version of the worldsheet in section
3. The calculation supports string formation in the intermediate
and strong coupling regimes. In section 4, the mean field method
is applied to the worldsheet lattice, where various details can
be treated more exactly than in the continuum approach. 
There the mean field approximation was identified with
a certain saddle-point evaluation of the worldsheet path
integral. Despite
its more solid starting point, this approach also suffers from
possible ambiguities in the choice of zeroth order
actions for the approximate evaluation. 
These ambiguities are partially resolved by
choosing a zeroth order action that has a transparent
formal continuum limit and agrees
well with an exact calculation of the contribution of
equally spaced solid lines in the zero coupling limit. The
outcome of the lattice calculation, which also results in
string formation in the strong coupling regime, is in
qualitative agreement with the results of section 3, although
there are some quantitative differences. The most notable of
these is the $3/2$ power dependence of the matter energy on the
field $\phi$ (Eq.~(\ref{smalleps})), as opposed to the linear dependence
given by (\ref{21}). We attribute this difference to the different
ways of handling the matter-ghost cancelation mentioned earlier,
and we are moderately encouraged that there is qualitative agreement.
We also recall the point made in Section 3
that neither the continuum nor the lattice methods 
adequately treat the so-called small black regions.
These correspond to fluctuations of the field $\phi$
over regions of the size of a small number of lattice sites. The
mean field approximation, which at least in the leading order,
treats $\phi$ as a constant over the whole worldsheet, neglects
these fluctuations. There is indication, for example from the
calculation of density of solid lines at the end of section 3,
that these small black regions may play an important role in
string formation. The identification made in Section 4 of the approximation
with a saddle-point evaluation provides a useful setting
to systematically explore the ramifications of these fluctuation effects.

Because of its various shortcomings mentioned above, we consider
the dynamical calculation presented in this article as a promising initial
attempt, not the final word. We hope that the line of approach
developed here will stimulate further progress in this important
problem of string formation in field theory.

\vskip.5cm

\noindent\underline{Acknowledgments:}
We would like to thank Jeff Greensite, Rajeesh Gopakumar, 
Igor Klebanov, Anatoly Konechny, and
Juan Maldacena for helpful conversations.
CBT benefited from the hospitality
of  the Institute for advanced Study at Princeton
and the Center for Theoretical Physics at MIT,
where part of this work was done. 
This work was supported in
part by the Monell Foundation, in part by
the Department of Energy under Grant No. DE-FG02-97ER-41029
and Contract No. DE-AC03-76F00098, and in part by the
National Science Foundation Grant PHY-0098840.

\appendix
\section{Determinants}
All of the Gaussian integrals we need in this paper are of the
generic form
\bea
{\cal I}&=&\int D{q}
\exp\left\{-\sum_{i,j}\left[A{({q}_{i}^j-{q}_{i}^{j-1})^2}
+B{({q}_{i+1}^j-{q}_{i}^{j})^2}\right]\right\}
\eea
These integrals can be done exactly using the
methods of \cite{gilest}. For simplicity we
impose periodic boundary conditions in $\sigma$ and
Dirichlet conditions in $\tau$, which describes
a closed string moving in a background of Ising spins.

The eigenvalues of the bilinear forms in the exponent are
well-known, so the integral which is proportional
to one over the square root of the determinant of the
bilinear form can be written as a double product, so
we find for two bosonic variables:
\bea
&&\hskip-.9cm{\cal I}^2=\left({\pi\over A}\right)^{MN}
\prod_{n=1}^N\prod_{l=0}^{M-1}\left(4\sin^2{n\pi\over2(N+1)}
+4{B\over A}\sin^2{l\pi\over M}\right)^{-1}
\eea
The analogous ghost integrals will be the reciprocal of
this expression (with of course different values for $A,B$.
We list some infinite products from \cite{gilest}:
\bea
\prod_{n=1}^N\left(4\sin^2\left({n\pi\over2(N+1)}
\right)-z\right)&=&{\sin(N+1)\kappa\over\sin\kappa},
\qquad z\equiv4\sin^2{\kappa\over2}\\
\prod_{l=1}^{M-1}\left(4\sin^2{l\pi\over M}-z\right)&=&{\sin^2M\lambda
\over\sin^2\lambda},\qquad z\equiv4\sin^2\lambda\\
\prod_{l=1}^{M-1}\left(2\sin{l\pi\over M}\right)=M,\qquad&&\qquad
\prod_{n=1}^N\left(2\sin{n\pi\over2(N+1)}\right)=\sqrt{N+1},
\eea
the last line being just the case $z=0$ of the first two.
Using these products, we then find
\bea
{\cal I}^2&=&{1\over{N+1}}\left({\pi\over A}\right)^{MN}
\prod_{l=1}^{M-1}\left({\sinh(N+1)\xi_l\over\sinh\xi_l}\right)^{-1},
\qquad \sinh^2{\xi_l\over2}={B\over A}\sin^2{l\pi\over M}
\eea
The ground state energy associated with this system of
two bosonic variables can
be read off by identifying the coefficient of $-N$ in $\ln{\cal M}$
as $N\to\infty$:
\bea
aE_G&=&-M\ln{2\pi a\over\beta m\phi^2}+2\sum_{l=1}^{M-1}\sinh^{-1}\left(
{\sqrt{\epsilon}\over {\phi}}\sin{l\pi\over M}\right)
\eea
where we note that $\sinh^{-1}(z)=\ln(z+\sqrt{1+z^2})$. So far everything
is exact. 
Now let's consider the continuum limit, $M\to\infty$. Note that
since $a,m\to0$ together in the continuum limit, it is
valid to keep $a/m$ fixed. We also keep $A,B$ fixed. 
The bulk contributions to this limit are straightforward:
\bea
{aE_G\over M}&\to&2\int_{0}^{1}dx\sinh^{-1}\left(
{\sqrt{B}\over\sqrt{A}}\sin{x\pi}\right)-
\ln{\pi\over A}=
{4\over\pi}{\rm Re}
\left\{
i{\rm Li}_2\left({\sqrt{B}\over i\sqrt{A}}\right)
\right\}
-\ln{\pi\over A}
\eea
We note the appearance of the dilogarithm or Spence function
${\rm Li}_2(x)={\rm dilog}(1-x)=\sum_{k=1}^\infty {x^n/n^2}$.

\section{Slowly Varying Mean Fields}
The identification of the mean field approximation with a
saddle point approximation to the path integral, enables
an easy extension to slowly varying fields. Indeed we have
already seen in the footnote to Eq.~\ref{contaction}, the
relatively simple extension of the effective action
for ghost$+$ matter fields to slowly varying fields.
Here we sketch the corresponding extension for the
spin system sum. This will give dynamical terms
to the mean field.

We first write out the saddle point equation for general
(non-uniform) $\phi, \lambda$.
\bea
\phi_{kl}={\sum_{s_i^j=\pm1}P_k^l
\left({\hat g}^\prime\right)^{(1-s_i^js_i^{j-1})/2}
e^{{-i\sum_{ij}\lambda_{ij} P_i^j}}\over\sum_{s_i^j=\pm1}
\left({\hat g}^\prime\right)^{(1-s_i^js_i^{j-1})/2}
e^{{-i\sum_{ij}\lambda_{ij} P_i^j}}}.
\eea
This equation implicitly determines $\lambda_{ij}(\phi)$ as a function
of all the $\phi_{ij}$. The equation is
intractable for general $\phi_{ij}$, but we have seen in the main text how to
explicitly solve the equation for uniform $\phi$, obtaining 
$\lambda_{ij}=i\kappa_0(\phi)$. We can get a
perturbative solution for slowly varying $\phi_{ij}=\phi_0+\delta\phi_{ij}$.
Actually, rather than $\lambda$ itself,
 we are more interested in expanding the effective action 
\bea
W_s(\phi,\lambda(\phi))&=&\sum_{ij}\phi_{ij}\lambda_{ij}-f(\lambda)\\
e^{-if(\lambda)}&\equiv& \sum_{s_i^j=\pm1}
\left({\hat g}^\prime\right)^{(1-s_i^js_i^{j-1})/2}
e^{{-i\sum_{ij}\lambda_{ij} P_i^j}}
\eea
about $\phi_0$. To expand to second order in $\delta\phi$, we only need
to compute first and second derivatives of $W_s$, which by virtue of the
fact that $W_s$ is a Legendre transform, are simply:
\bea
{\partial W_s\over\partial \phi_{ij}}&=&\lambda_{ij}(\phi_0)\\
{\partial^2 W_s\over\partial\phi_{mn}\partial\phi_{ij}}&=&{\partial\lambda_{ij}
\over\partial\phi_{mn}}\bigg|_{\phi=\phi_0}.
\eea
Then the expansion to quadratic order reads
\bea
W_s(\phi,\lambda(\phi))=W_s(\phi_0,\lambda(\phi_0))+\sum_{ij}
\delta\phi_{ij}i\kappa_0(\phi_0)
+{1\over2}\sum_{ij,mn}\delta\phi_{ij}\delta\phi_{mn}{\partial\lambda_{ij}
\over\partial\phi_{mn}}\bigg|_{\phi=\phi_0}+O(\delta\phi^3).
\eea
The matrix $\partial\lambda/\partial\phi$ is the inverse of the
matrix $\partial\phi/\partial\lambda$, which can be related to
spin correlators by
\bea
{\partial\phi_{ij}\over\partial\lambda_{mn}}
&=&-i[\langle P_i^jP_m^n\rangle-\langle P_i^j\rangle\langle P_m^n\rangle]\\
\langle\Omega\rangle&\equiv&
{\sum_{s_i^j=\pm1}\Omega\left({\hat g}^\prime\right)^{(1-s_i^js_i^{j-1})/2}
e^{{\sum_{ij}\kappa_0(\phi_0) P_i^j}}\over\sum_{s_i^j=\pm1}
\left({\hat g}^\prime\right)^{(1-s_i^js_i^{j-1})/2}
e^{{\sum_{ij}\kappa_0(\phi_0)P_i^j}}}.
\eea
Since these spin correlators are all in the presence of a uniform
background field $\lambda$, they may be explicitly evaluated in
terms of the eigenvalues of the spin transfer matrix introduced in
Section 4.
\bea
\langle P_i^jP_m^n\rangle=\langle P_i^j\rangle\langle P_m^n\rangle
+\delta_{im}\left({t_-\over t_+}\right)^{|j-n|}{{\hat g}^{\prime2}
\over4({\hat g}^{\prime2}+\sinh^2(\kappa_0/2))}
\eea
so that
\bea
{\partial\phi_{ij}\over\partial\lambda_{mn}}
&=&-i\delta_{im}\left({t_-\over t_+}\right)^{|j-n|}{{\hat g}^{\prime2}
\over4({\hat g}^{\prime2}+\sinh^2(\kappa_0/2))}.
\eea

The matrix $M_{jn}=\rho^{|j-n|}$ can be inverted by defining
$w_j=\sum_n\rho^{|j-n|}v_n$ and proving the recursion relation
\bea
w_{j+1}+w_{j-1}=(\rho+1/\rho)w_j+(\rho-1/\rho)v_j,
\eea
from which it follows that
\bea
(M^{-1})_{jn}&=&
{\rho\over \rho^2-1}(\delta_{j+1,n}+\delta_{j-1,n}-2\delta_{jn})
-{\rho-1\over \rho+1}\delta_{jn}\nonumber\\
&=&-{(1-{\hat g}^{\prime2})(\delta_{j+1,n}+\delta_{j-1,n}-2\delta_{jn})
\over4\cosh(\kappa/2)\sqrt{{\hat g}^{\prime2}
+\sinh^2(\kappa/2)}}
+\delta_{jn}{\sqrt{{\hat g}^{\prime2}
+\sinh^2(\kappa/2)}\over\cosh(\kappa/2)}.
\eea
Plugging these results into the quadratic approximation
to the spin system effective action then gives
\bea
iW_s(\phi,\lambda(\phi))&\approx&iW_s(\phi_0,i\kappa_0(\phi_0))-\sum_{ij}
\delta\phi_{ij}\kappa_0(\phi_0)
-{1\over2}\sum_{ij}\left[Z(\delta\phi_{ij}-\delta\phi_{i,j-1})^2
+\mu^2\delta\phi^2_{ij}\right]\\
Z&\equiv&{(1-{\hat g}^{\prime2})\sqrt{{\hat g}^{\prime2}
+\sinh^2(\kappa/2)}\over{\hat g}^{\prime2}\cosh(\kappa/2)},\qquad
\mu^2\equiv4{({\hat g}^{\prime2}
+\sinh^2(\kappa/2))^{3/2}\over{\hat g}^{\prime2}\cosh(\kappa/2)}.
\eea
This result gives the spin system effective action for $\phi_{ij}=\phi_0
+\delta\phi_{ij}$ to quadratic order in $\delta\phi$. Formally
$\phi_0$ can be chosen to be anything, but to minimize corrections
to the slowly varying mean field
approximation,
it should be chosen so that $\sum_{ij}\delta\phi_{ij}=0$. In the uniform
mean field approximation studied in the main text, this means
simply $\phi_{ij}=\phi=\phi_0$.

It is instructive to combine the work of this appendix with that
in the footnote to Eq.~\ref{contaction} to give the formal
continuum limit of the total effective action to order $\delta\phi^2$:
\bea
S&=&-iW_s(\phi_0,i\kappa_0(\phi_0))+ 
\int d\tau d\sigma \left({1\over2}{\bf q}^{\prime2}
+{a^2\phi\over 2m^2\epsilon}{\dot{\bf q}}^2-(1-\phi^2)^2b^\prime c^\prime
-{a^2\phi\over m^2\epsilon}{\dot b}{\dot c}\right.\nonumber\\
&&\left.\hskip3cm
-{1\over2}(\phi^{\prime\prime}+2\phi^{\prime2})bc+{aZ\over2m}{\dot\phi}^2
+{\mu^2\over2am}(\phi-\phi_0)^2\right)+O(\phi-\phi_0)^3.
\eea
This equation expresses the mean field dynamics for our system in
a form tantalizingly close to the outcome of the Maldacena
conjecture \cite{maldacena}. The mean field has emerged as
a dynamical Liouville-like worldsheet field that enters the
worldsheet action in a way strongly reminiscent of the way
the radial AdS coordinate enters the worldsheet action
for the ``string'' description of ${\cal N}=4$ super-Yang-Mills
theory.


\begin{thebibliography}{1}
\bibitem{bardakcit}
K.~Bardakci and C.~B.~Thorn,
Nucl.\ Phys.\ B {\bf 626} (2002) 287
[arXiv:hep-th/0110301].

\bibitem{maldacena}
J. M. Maldacena, {\sl Adv. Theor. Math. Phys.} {\bf 2} (1998) 231-252,
  hep-th/9711200.

\bibitem{thornsheet}
C.~B.~Thorn,
arXiv:hep-th/0203167.

\bibitem{gilest}
R. Giles and C. B. Thorn, {\sl Phys. Rev.} {\bf D16} (1977) 366.

\bibitem{orland}
P.~Orland,
Nucl.\ Phys.\ B {\bf 278} (1986) 790.


\bibitem{dalleyk}
S.~Dalley and I.~R.~Klebanov,
Phys.\ Lett.\ B {\bf 298} (1993) 79
[arXiv:hep-th/9207065].


\bibitem{greensiteh}
J. Greensite and M. B. Halpern, Nucl.\ Phys.\ B {\bf 242} (1984) 
167. 


\bibitem{thooftfiniteradius}
G.~'t Hooft,
Phys.\ Lett.\ B {\bf 119} (1982) 369,
Commun.\ Math.\ Phys.\  {\bf 88} (1983) 1.

\end{thebibliography}
\end{document}